\documentclass[aps,showpacs,twocolumn,superscriptaddress,prb,floatfix, 10pt, longbibliography]{revtex4-2}

\usepackage[utf8]{inputenc}
\usepackage{amsfonts}
\usepackage{amsmath}
\usepackage{amssymb}
\usepackage{bbm}
\usepackage{amsthm}
\usepackage{empheq}
\usepackage{enumitem}
\usepackage{verbatim}
\usepackage{color}
\usepackage{subfigure}
\usepackage[breaklinks]{hyperref}
\usepackage{graphicx}
\usepackage{times}

\begin{document}

\title{Hedgehog lattices induced by chiral spin interactions}
\author{Ryan Mays}
\affiliation{Department of Physics and Astronomy, George Mason University, Fairfax, VA 22030, USA}

\author{Predrag Nikoli\'c}
\affiliation{Department of Physics and Astronomy, George Mason University, Fairfax, VA 22030, USA}
\affiliation{Institute for Quantum Matter and Department of Physics and Astronomy, Johns Hopkins University, Baltimore, MD}

\date{\today}

\begin{abstract}

We analyze a classical Heisenberg spin model on the simple cubic lattice which is invariant under time reversal and contains multiple chiral spin interactions. The modelled dynamics is appropriate either for local moments coupled to itinerant Weyl electrons, or localized electrons with a strong spin-orbit coupling that would produce a Weyl spectrum away from half filling. Using a Monte Carlo method, we find a robust $4Q$ bipartite lattice of hedgehogs and antihedgehogs which melts through a first order phase transition at a critical temperature in certain segments of the phase diagram. The density of hedgehogs is a non-linear function of the Dzyaloshinskii-Moriya interaction, and a linear function of the multiple-spin chiral interaction which plays the fundamental role of a ``magnetic flux'' or a hedgehog chemical potential. These findings are related to the observations of hedgehog lattices in MnGe, MnSi$_{1-x}$Ge$_x$ and SrFeO$_3$, and indirectly support the possible existence of incompressible quantum-disordered hedgehog liquids.

\end{abstract}

\maketitle

\section{Introduction}

Topological defects play an instrumental role in many unconventional states of matter. The best-known example are Abrikosov vortices in superconductors \cite{Abrikosov1957}. A simple Landau-Ginzburg theory explains how quantized vortices arise in the external magnetic field, and why they order into a lattice in type-2 superconductors. Thermally disordered vortices may be responsible for the puzzling Nernst effect in the underdoped pseudogap state of cuprates \cite{Ong2001, Wang2006}. There is also a close connection to quantum Hall physics: quantum melting of an Abrikosov lattice can produce a fractional quantum Hall liquid \cite{Cooper1999a, Cooper2001, Cooper2008} whose elementary constituents are bound states of a quantized vortex and a fraction of fundamental charge. This phenomenology is quite general. It applies to any system whose degrees of freedom support singular topological defects \cite{Nikolic2019}. Three-dimensional magnets can form skyrmion line defects, as well as point defects known as hedgehogs or monopoles. Only hedgehogs enjoy true topological protection in isotropic magnets. It has been understood for some time that skyrmions and hedgehogs can be induced by chiral spin interactions, most notably the Dzyaloshinskii-Moriya interaction. However, it is seldom emphasized that the direct spin analogue of the magnetic field for Abrikosov vortices are the three-spin and four-spin chiral couplings \cite{Nikolic2019b}. Recent experiments in non-centrosymmetric magnets MnGe and MnSi$_{1-x}$Ge$_x$ have revealed stable hedgehog lattices with different structures \cite{Tokura2011, Kanazawa2012, Tokura2013b, Tokura2015, Fujishiro2019, Tokura2020, Tokura2021, Nakajima2023}, and similar textures have also been found in the centrosymmetric material SrFeO$_3$ \cite{Tokura2020b}. These states may be amenable to external control by an applied magnetic field or current. The spacing between hedgehogs is fairly small in these materials, so a hypothetical similar system with spins of low magnitude could be close to a quantum melting regime, perhaps an exotic fractionalized phase.

Since many questions about topological magnetism are still open, these insights from experiments and theory motivate us to study simple models of chiral spins and analyze the dynamics of hedgehog lattices. Here we consider an effective spin model with Heisenberg, Kitaev-like, Dzyaloshinskii-Moriya and multiple-spin chiral interactions which can arise by two different microscopic mechanisms. One is the effect of itinerant Weyl electrons on coexisting local moments; the induced RKKY interactions of the mentioned types appear at various orders of perturbation theory in the Kondo/Hund coupling \cite{Nikolic2020a}. The range across which these interactions are significant is limited by the ultra-violet cut-off momentum of the Weyl spectrum. Another mechanism is the localization of electrons at half-filling due to Coulomb interactions, provided that the electrons microscopically have a spin-orbit coupling which could produce a Weyl spectrum away from half filling \cite{Nikolic2019b}. In this case, the spin interactions would be short-ranged; the resulting antiferromagnetic exchange is not an obstacle for the formation of hedgehogs in the ``staggered spin'' order parameter.

We are primarily interested in the broken symmetry phases and the disordering phase transition, so we employ a classical Monte Carlo technique to study the spin dynamics. As anticipated from Landau-Ginzburg-type arguments \cite{Nikolic2019b}, we find that the chiral interactions of our model stabilize a robust lattice of hedgehogs and antihedgehogs at low temperatures. We observe only the simplest $4Q$ lattice in the studied portion of the phase diagram, possibly enhanced by the cubic structure of the underlying microscopic lattice and a typically short distance between the defects. The phase transition at a finite critical temperature appears to be of first order in a certain portion of the phase diagram, which is consistent with a thermal melting of the hedgehog lattice. The temperature dependence of the extracted specific heat bears some similarity to the experimental observations \cite{Bauer2013} of skyrmion lattice melting in MnSi. A first order transition is also promising for the prospects of stabilizing topologically ordered chiral quantum spin liquids in three dimensions \cite{Nikolic2019, Nikolic2019b}.

Our approach to hedgehog lattices supplements previous theoretical studies \cite{Binz2006, Han2011, Kanazawa2016, Nagaosa2016, Han2016, Bornemann2019, Motome2019, Motome2019b, Kawamura2020, Motome2020, Motome2021, Motome2021a, Motome2022a, Motome2022b, Motome2022c, Okumura2022, Yudin2022, Mochizuki2024} in several ways. The present analysis is most similar to the early Monte Carlo work \cite{Han2016} where a hedgehog lattice was found only at finite temperatures and in external magnetic fields, in a model with nearest and second neighbor Heisenberg and Dzyaloshinskii-Moriya interactions. We extend this model with higher-order chiral interactions that are expected to occur in complex magnets \cite{Blugel2020, Nikolic2019b}, and observe a robust hedgehog lattice down to zero temperature and in zero field, resulting from spontaneous symmetry breaking. Importantly, Monte Carlo treats hedgehogs as mobile topological defects whose positional fluctuations may enhance or destabilize different phases and lattice arrangements. In contrast, the typical theoretical approach in recent years has been to represent a hedgehog lattice as a ``superposition'' of several spin helices at different wavevectors, and then minimize the (free) energy under variations of the ``superposition'' parameters. This procedure is very powerful for extracting the phase diagram of this complex system, but does not take into account the positional fluctuations of hedgehogs. Positional flexibility is certainly important for stabilizing defect lattices in real materials where disorder cannot be neglected. Furthermore, our model contains multiple chiral interactions which are responsible for nucleating hedgehogs, while only the Dzyaloshinskii-Moriya interaction has been commonly studied before. Our interactions have a short range, while other works typically simplify the analysis by assuming an infinite range \cite{Motome2019, Motome2019b, Motome2020, Motome2021, Motome2022a, Motome2022b, Motome2022c, Okumura2022, Yudin2022}, motivated by the itinerant electron origin; an infinite range, however, may skew the stability of various phases. In this regard, our calculations are complementary to the previous literature, and explore the hedgehog lattice physics in a broader context of chiral magnet models.

This paper is organized as follows. Section \ref{secModel} introduces the model we study and motivates it by its physical features expected from the spin-orbit coupling. Section \ref{secMonteCarlo} describes our Monte Carlo simulations of dynamics in this model, and discusses the hedgehog lattice we find at low temperatures. The properties of the hedgehog lattice which depend on the chiral interactions, as well as its phase transition to a disordered phase are discussed. All conclusions are summarized in Section \ref{secConclusions}.

We often use index notation in this paper where lower Greek indices $\mu,\nu,\dots\in\lbrace x,y,z\rbrace$ represent spatial directions on the lattice and upper indices $a,b,\dots\in\lbrace x,y,z\rbrace$ denote spin vector components. Repeated indices are summed over by Einstein's convention in all formulas where scalar vector/tensor components are explicitly written. Levi-Civita tensor is denoted by $\epsilon_{\mu\nu\lambda}$ or $\epsilon^{abc}$. Sometimes we use boldcase symbols for spin vectors.

\section{The model}\label{secModel}

The spin model we study on the cubic lattice contains several types of spin interactions
\begin{equation}\label{Hamiltonian}
H=H_{\textrm{H}}+H_{\textrm{K}}+H_{\textrm{Z}}+H_{\textrm{DM}}+H_{\textrm{c}}+H_{\textrm{C}}+H_{\chi} \ .
\end{equation}
Its basic ingredient is the Heisenberg interaction
\begin{equation}\label{HeisenbergInteraction}
H_{\textrm{H}} = -\sum_{ij} J_{ij}\, {\bf S}_{i}\cdot{\bf S}_{j}
\end{equation}
between pairs of spins on nearby lattice sites $i,j$. For simplicity, we focus on the nearest-neighbor interactions only and keep $J$ positive, thus aligning the spins ferromagnetically. Since the lattice is bipartite, antiferromagnetic $J<0$ interactions are effectively included in our calculations through the transformation ${\bf S}_i\to-{\bf S}_i$ on all sites of one sub-lattice. Spin RKKY interactions induced by Weyl electrons also include a Kitaev-Ising coupling \cite{Nikolic2020a}
\begin{equation}\label{KitaevInteraction}
H_{\textrm{K}} = -\sum_{i}\Bigl(K_{x}^{\phantom{x}}S_{i}^{x}S_{i+\hat{{\bf x}}}^{x}+K_{y}^{\phantom{x}}S_{i}^{y}S_{i+\hat{{\bf y}}}^{y}+K_{z}^{\phantom{x}}S_{i}^{z}S_{i+\hat{{\bf z}}}^{z}\Bigr) \ ,
\end{equation}
here taken on nearest-neighbor bonds and kept isotropic, $K_x=K_y=K_z\equiv K$. Our calculations do not include multiple-spin interactions that preserve the inversion symmetry, such as the frequently considered biquadratic coupling \cite{Motome2019, Motome2019b, Motome2020, Motome2021, Motome2022a, Motome2022b, Okumura2022, Yudin2022}.

Apart from the Zeeman coupling $H_{\textrm{Z}} = -\mu_0 \sum_i {\bf B}\cdot {\bf S}_i$, the remaining spin interactions are chiral, aimed at stabilizing topologically non-trivial magnetic states. The strongest among these is usually Dzyaloshinskii-Moriya (DM)
\begin{equation}\label{DMInteraction}
H_{\textrm{DM}}=\sum_{i}\sum_{\mu}{\bf D}_{\mu}\cdot({\bf S}_{i}\times{\bf S}_{i+\hat{\boldsymbol{\mu}}}) \to \int d^3x\, A_{\mu}^{a}J_{\mu}^{a} \ .
\end{equation}
As a template, we consider the DM vector ${\bf D}_\mu = D \hat{\boldsymbol{\mu}}$ parallel to lattice bonds $\hat{\boldsymbol{\mu}}\in\lbrace\hat{{\bf x}},\hat{{\bf y}},\hat{{\bf z}}\rbrace$, which is induced by isotropic Weyl electrons whenever lattice symmetries permit \footnote{The DM vector is perpendicular to the lattice bonds which are bisected by mirror symmetry planes. However, the long range of the RKKY DM interactions and the reduced symmetries of the chiral magnetically ordered phases conspire to blunt the effect of these symmetry restrictions.}, as a component of RKKY interactions \cite{Nikolic2020a}. The same interaction occurs among localized spin-orbit-coupled electrons as well \cite{Nikolic2019b}. The continuum limit of the DM interaction, shown behind the arrow, is simply the minimal coupling of the spin current $J_{\mu}^{a}=\epsilon^{abc}S^{b}\partial_{\mu}S^{c}$ to a static background SU(2) gauge field $A_\mu^a \equiv D_\mu^a$ given by the DM vector itself. This provides a significant intuition about spin dynamics, based on the fairly analogous electromagnetism of the charge current coupled to a U(1) gauge field as $j_\mu A_\mu$. The bond-alignment takes form $D_{\mu}^{a}=D\delta_{\mu}^{a}$ in terms of spatial $\mu$ and spin $a$ indices; we confirm numerically that this stimulates the appearance of hedgehogs, as anticipated with a Landau-Ginzburg-type argument \cite{Nikolic2019b}.

The next two Hamiltonian terms are the multiple-spin chiral interactions induced by effective magnetic fluxes,
\begin{eqnarray}\label{ChiralInteraction}
H_{\textrm{c}} &=& \sum_{i}\sum_{\mu\nu}\phi_{\mu\nu}\,{\bf S}_{i}\cdot({\bf S}_{i+\hat{\boldsymbol{\mu}}}\times{\bf S}_{i+\hat{\boldsymbol{\nu}}}) \\
H_{\textrm{C}} &=& \sum_{i}\sum_{\mu\nu}(\boldsymbol{\Phi}_{\mu\nu}\cdot{\bf S}_{i})\,{\bf S}_{i}\cdot({\bf S}_{i+\hat{\boldsymbol{\mu}}}\times{\bf S}_{i+\hat{\boldsymbol{\nu}}}) \ . \nonumber
\end{eqnarray}
The three-spin interaction $H_c$ is the well-known coupling \cite{Chitra1995} induced in an external magnetic field ${\bf B}$ whose flux through a lattice plaquette $\mu\nu$ is given by $\phi_{\mu\nu}$. It stimulates the appearance of skyrmions by providing an effective ``chemical potential'' to the scalar spin chirality. This is a handle for the experimental manipulation of topological magnetic states, but we do not consider it in this study because we are primarily interested in the states which break the time-reversal symmetry spontaneously. Seldom discussed is its SU(2) counterpart $H_{\textrm{C}}$ obtained from the spin-orbit coupling \cite{Blugel2020}. This arises at the third order of perturbation theory for localized electrons at half-filling \cite{Nikolic2019b}, just like $H_{\textrm{c}}$, and is also generated by the presence of itinerant Weyl electrons. $H_{\textrm{C}}$ does not explicitly break the time-reversal symmetry, but becomes activated only in the presence of local magnetization. Specifically, the spin-orbit coupling behind the isotropic Weyl spectrum produces $\Phi_{\mu\nu}^{a}=\Phi\epsilon_{a\mu\nu}$. Such a coupling is on par with the biquadratic interaction $({\bf S}_i\cdot {\bf S}_j)^2$, which is sought as a stabilizer of hedgehog states in centrosymmetric materials. We confirm numerically here that $H_{\textrm{C}}$ generates hedgehogs, as expected from the Landau-Ginzburg-type argument \cite{Nikolic2019b}. We implement this interaction in Monte Carlo by sampling the spins on the vertices of 12 isosceles right-angled triangles for each lattice site $i$ (the catheti of the triangles are nearest-neighbor bonds). Together, $H_{\textrm{c}}$ and $H_{\textrm{C}}$ represent the coupling of the rank-2 ``skyrmion'' current $J_{\mu\nu}=\epsilon^{abc}S^{a}(\partial_{\mu}S^{b})(\partial_{\nu}S^{c})$ to the rank-2 gauge field $A_{\mu\nu}^{\phantom{x}}=\phi_{\mu\nu}^{\phantom{x}}+\Phi_{\mu\nu}^{a}S^{a}$ found in field-theoretical descriptions of chiral magnets \cite{Nikolic2019b}. In fact, this rank-2 gauge field acts as the ``chemical potential'' for hedgehogs exactly the same way as the ordinary magnetic field does for vortices in superconductors.

The last term in (\ref{Hamiltonian}) is the chiral energy defined on the site triplets just like the previous interaction:
\begin{equation}\label{ChiralEnergy}
H_{\chi}=C_{\chi}\sum_{i}\sum_{\mu\nu}\Bigl\lbrack{\bf S}_{i}\cdot({\bf S}_{i+\hat{\boldsymbol{\mu}}}\times{\bf S}_{i+\hat{\boldsymbol{\nu}}})\Bigr\rbrack^{2} \ .
\end{equation}
This interactions appears at higher orders of perturbation theory in microscopic models, so the coupling constant $C_\chi$ is expected to be small. We neglect this interaction in the present study, but note that it represents an effective Maxwell term for the non-Abelian part of the rank-2 gauge field, generated due to spin fluctuations.

Our goal behind this Hamiltonian construction is to explore the topological analogies between the dynamics of spins and superconductors. An SU(2) magnetic order parameter supports singular topological defects in three dimensions, characterized by an integer-valued invariant of the $\pi_2(S^2)$ homotopy group. Different names, monopoles and hedgehogs, are sometimes used to emphasize their visual appearance, but here we will focus only on their topologically protected content as we use the term hedgehogs generically. Just as the $\pi_1(S^1)$ vortices of superconductors, the hedgehogs can be generated by the flux of an appropriate gauge field. The $\pi_2(S^2)$ homotopy requires a rank-2 gauge field for the formulation of its integer-valued invariant as a flux quantum, and the relationship between the hedgehogs and the rank-1 gauge field, i.e. the DM interaction, is indirect \cite{Nikolic2019}. One important difference, however, is that the microscopic material conditions stimulate lattices of hedgehogs and antihedgehogs, instead of the topological defects that all carry the same topological charge as in superconductors.

Itinerant Weyl electrons will tend to produce ``ferromagnetic'' Heisenberg and Kitaev-Ising interactions, although spatially modulated in accordance to the momentum-space separation between Weyl nodes \cite{Nikolic2020a}. However, localized electrons at half-filling have antiferromagnetic interactions. Even though we simulate ferromagnetic interactions for simplicity, antiferromagnetism obtains by a spin reversal on all sites of one sub-lattice. This changes the sign of the nearest-neighbor DM coupling, which is not a significant qualitative change, and does not effect the SU(2) chiral interaction $H_{\textrm{C}}$ at all. The three-spin chiral interaction $H_{\textrm{c}}$ changes sign, but we do not simulate this here; it does invert the chiral response to the external magnetic field, though, relative to ferromagnets. For these reasons, the results we obtain are qualitatively applicable both to ferromagnets and antiferromagnets.

\section{Monte Carlo simulations}\label{secMonteCarlo}

We simulate the model (\ref{Hamiltonian}) of classical spins on the cubic lattice using the standard Monte Carlo Metropolis-Hastings algorithm. In order to characterize the topologically non-trivial magnetic orders, we measure the local magnetization $\langle S_i \rangle$ on every lattice site, and the spin chirality on the smallest triangles of lattice sites
\begin{equation}\label{SpinChirality}
\langle\chi_{ijk}\rangle = \langle {\bf S}_i\cdot ({\bf S}_j \times {\bf S}_k) \rangle
\end{equation}
as functions of position on the lattice. Note that the chirality is a scalar under spin rotations, but a pseudovector $\boldsymbol{\chi}$ under spatial point-group transformations since it can be viewed as a ``vector'' perpendicular to the oriented plane $ijk$. When the measurements of $\langle S_i \rangle$ and $\langle\chi_{ijk}\rangle$ provide evidence of topological structures, we also explicitly compute the hedgehog $\pi_2(S^2)$ topological charges on the cubic unit-cells of the lattice by first interpolating the eight spins on the cube's vertices into a continuous configuration ${\bf S}({\bf r})$ inside the cell, and then evaluating
\begin{equation}\label{TopCharge}
\mathcal{N} = \frac{1}{4\pi} \oint d^2 x\, \hat{n}_\mu \epsilon_{\mu\nu\lambda} {\bf S}\cdot (\partial_\nu {\bf S} \times \partial_\lambda {\bf S})
\end{equation}
over the cell's boundary, where $\hat{\bf n}$ is the unit vector locally perpendicular to the boundary. This interpolation method is fairly reliable, although ambiguous regarding the possible appearance of hedgehog-antihedgehog pairs inside a single cell. Skyrmions are seen as regions of large chirality concentrated into tubes, while hedgehogs act as point sources of chirality, analogous to electric charges as sources of electric field.

Monte Carlo runs presented here employ the simplest Metropolis-Hastings algorithm with updates that rotate a single lattice spin in an unbiased random direction. Since the spin textures form complex topological textures, it typically takes a large number of Monte Carlo iterations (update attempts) to reach equilibrium. We often simulate annealing in order to increase the chance of discovering the true equilibrium states. During annealing, the temperature is ramped from an initial high value down to the desired final value, exponentially as a function of the Monte Carlo time (iteration count). Measurements of the spin configuration are taken only after the system reaches an apparent equilibrium (energy fluctuations stop evolving), and many iterations apart to ensure that every spin had a chance to receive multiple updates. An alternative protocol used in some runs involves ``natural'' annealing by taking fixed-temperature measurements between small temperature changes in gradual up/down temperature sweeps; this facilitates the discovery of 1st order phase transitions. In most cases, we collected multiple data sets for same model parameters and compared their internal energy in order to minimize the likelihood of being sidetracked by metastable states. The landscape of metastable states was explored to a certain extent during this process.

\begin{figure*}[!t]
\centering
    \subfigure[{}]{\includegraphics[height=1.9in]{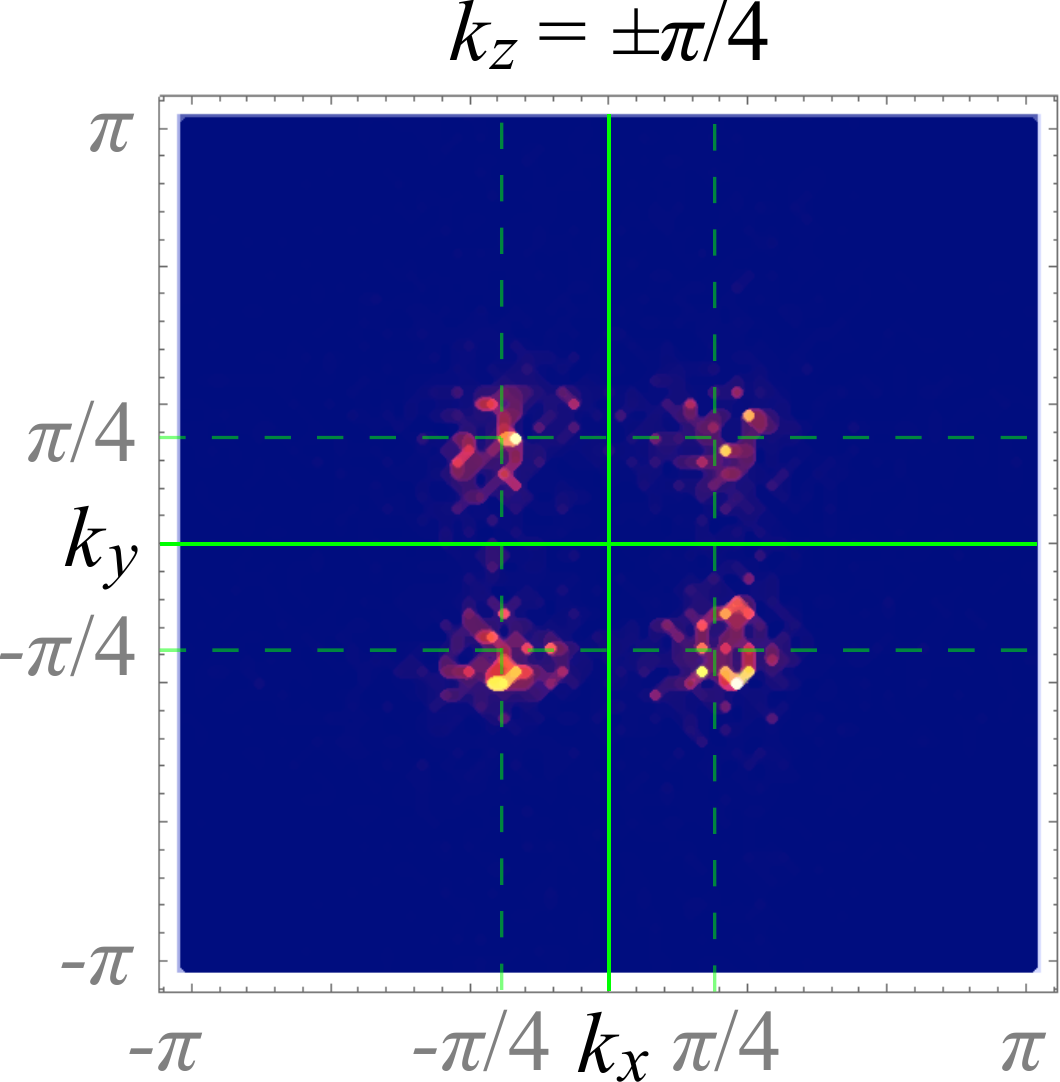}}\hspace{0.2in}
    \subfigure[{}]{\includegraphics[height=1.9in]{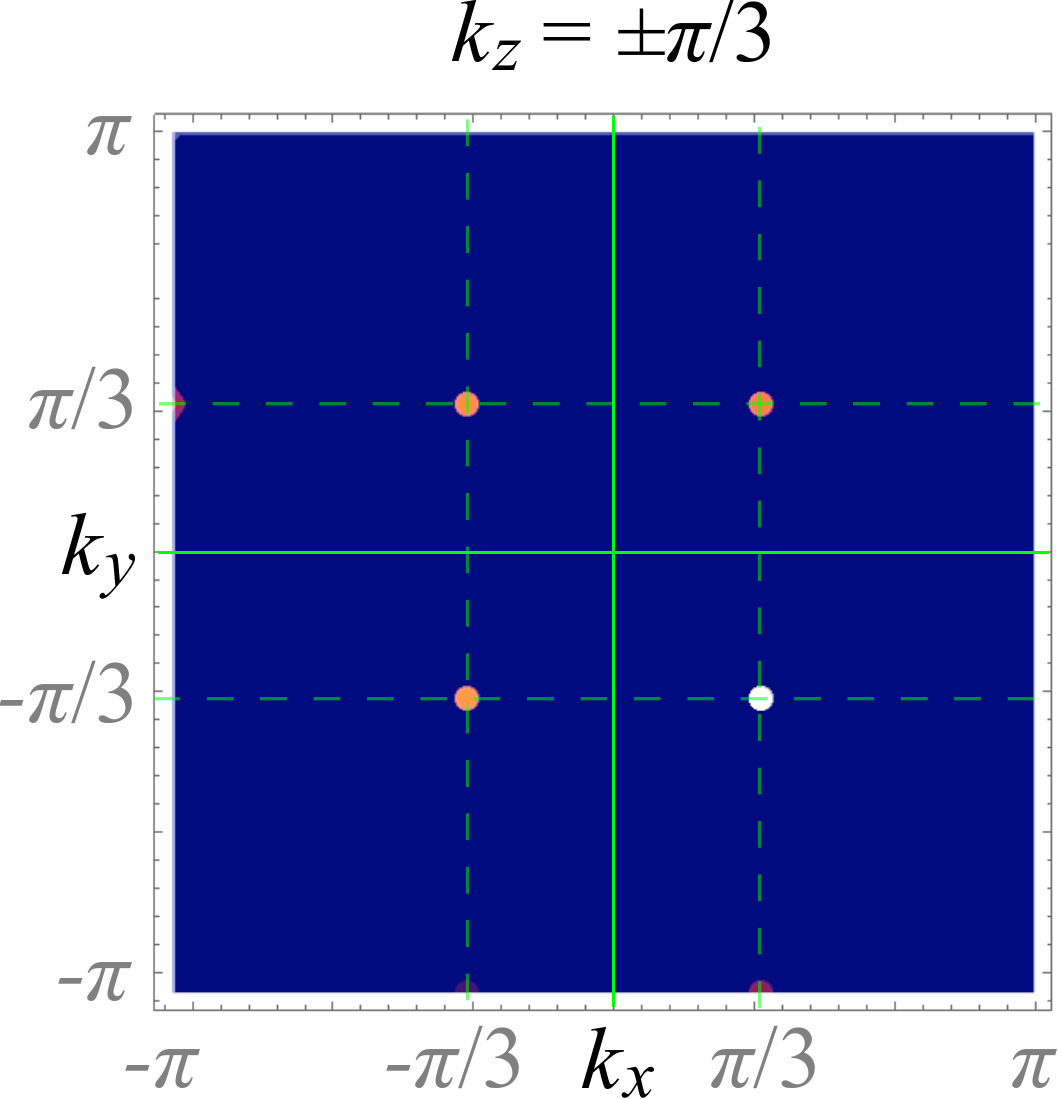}}\hspace{0.2in}
    \subfigure[{}]{\includegraphics[height=1.9in]{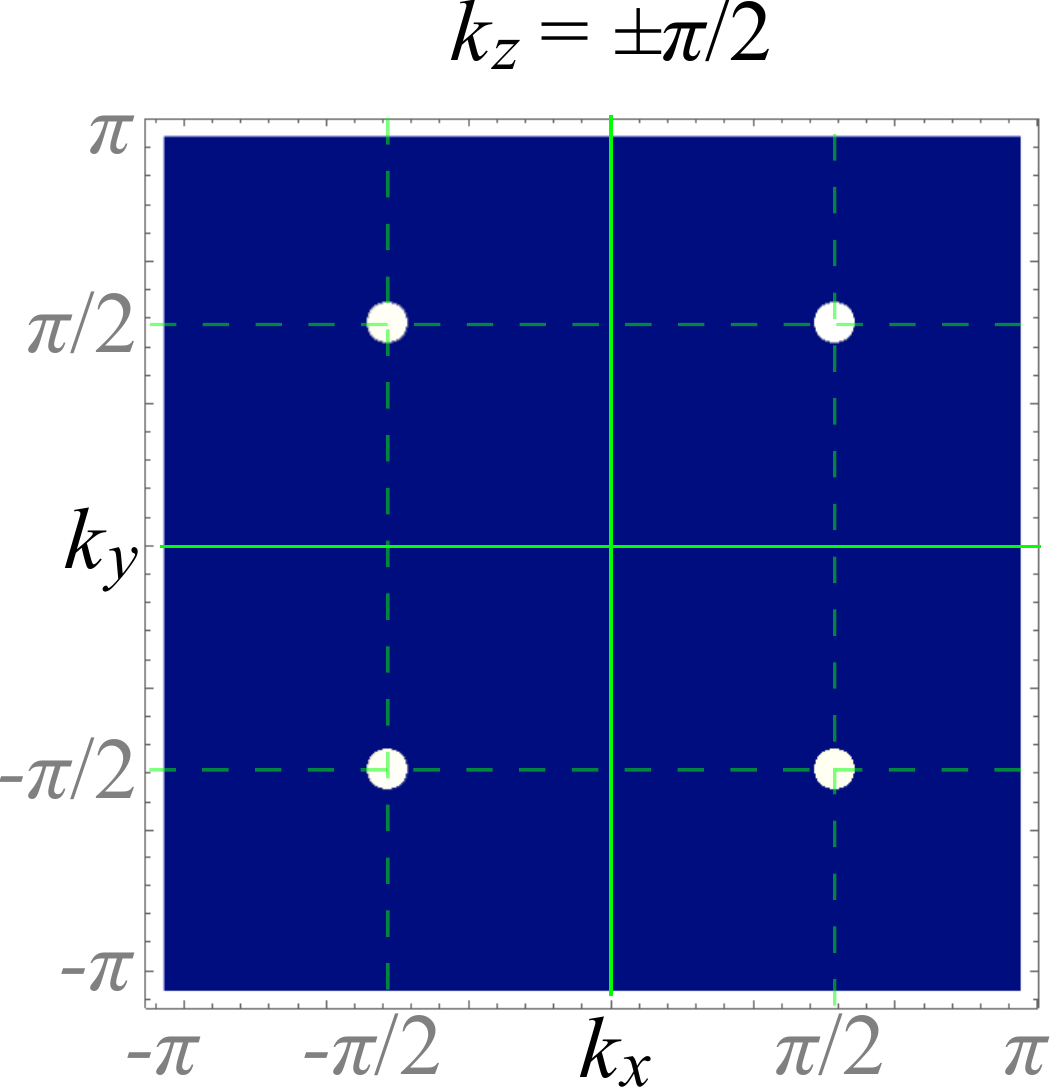}}
    \caption{\label{ChiralityPlots}The fast Fourier transform (FFT) amplitude of the magnitude $|\boldsymbol{\chi}|$ of the spin chirality pseudovector (\ref{SpinChirality}) in three representative hedgehog lattices: (a) $(D,\Phi)=(1.0,0.4)$, $N=75^3$ at $T=0$, (b) $(D,\Phi)=(1,1)$, $N=42^3$ at $T=0$ and (c) at $(D,\Phi)=(1,1)$, $N=24^3$ at $T=1.5$ ($J=1$, $K=C=0$ in all cases). The peaks of the chirality FFT generally have a roughly spherical shape and sit at eight vertices of a cube in momentum space. The simulation parameters were: (a) $800$ measurements in $2\times10^8$ iterations after annealing, (b) $2000$ measurements in $8.75\times10^8$ iterations after annealing, and (c) $10^3$ measurements in $2.5\times 10^8$ iterations with $32\%$ acceptance of single-spin updates. Note that the state (c) is metastable at the given model parameters, but becomes stable at larger values of $D$ or $\Phi$.}
\end{figure*}

\begin{figure}[!t]
    \subfigure[{}]{\includegraphics[width=1.7in]{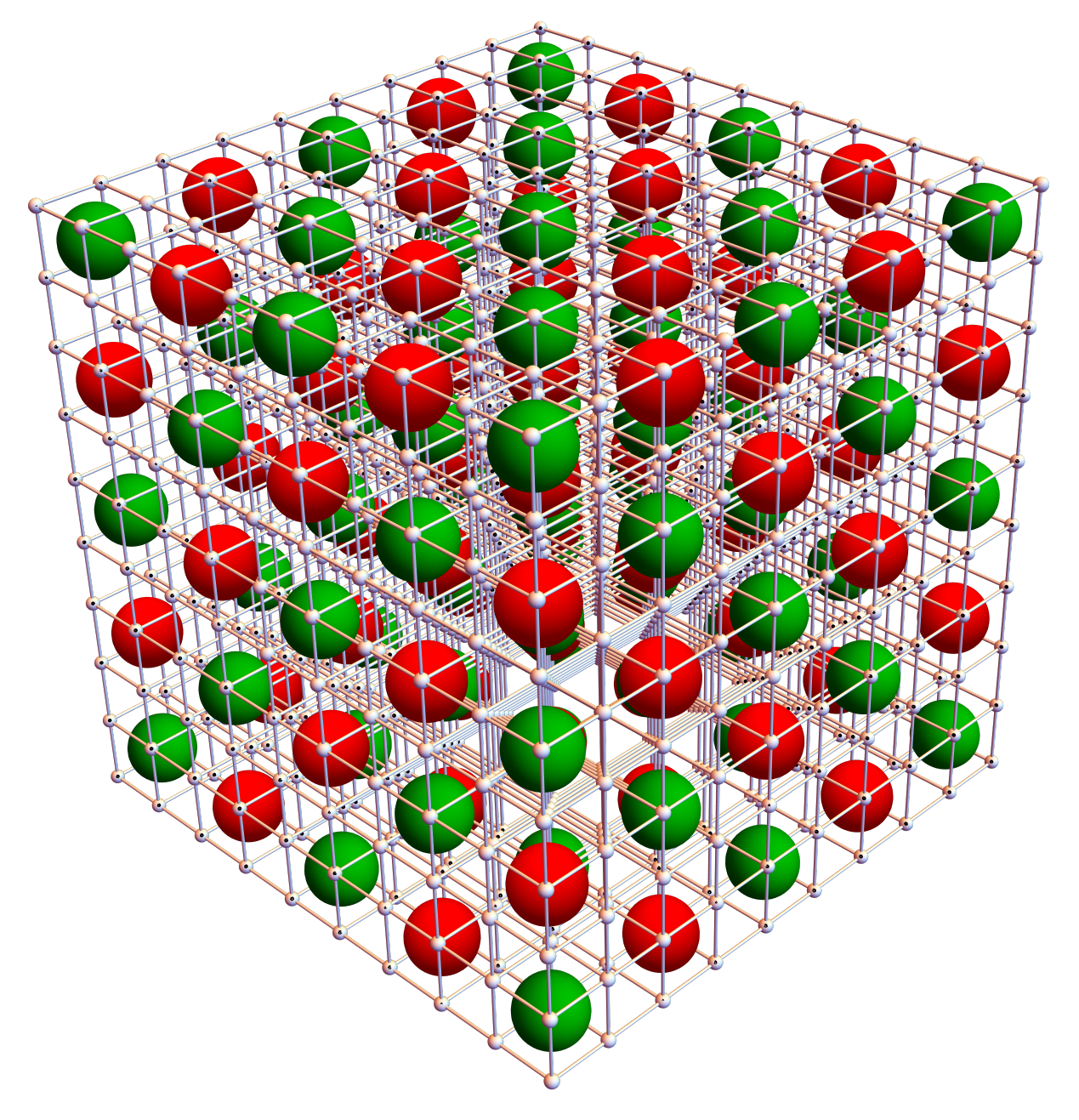}}
    \subfigure[{}]{\includegraphics[width=1.6in]{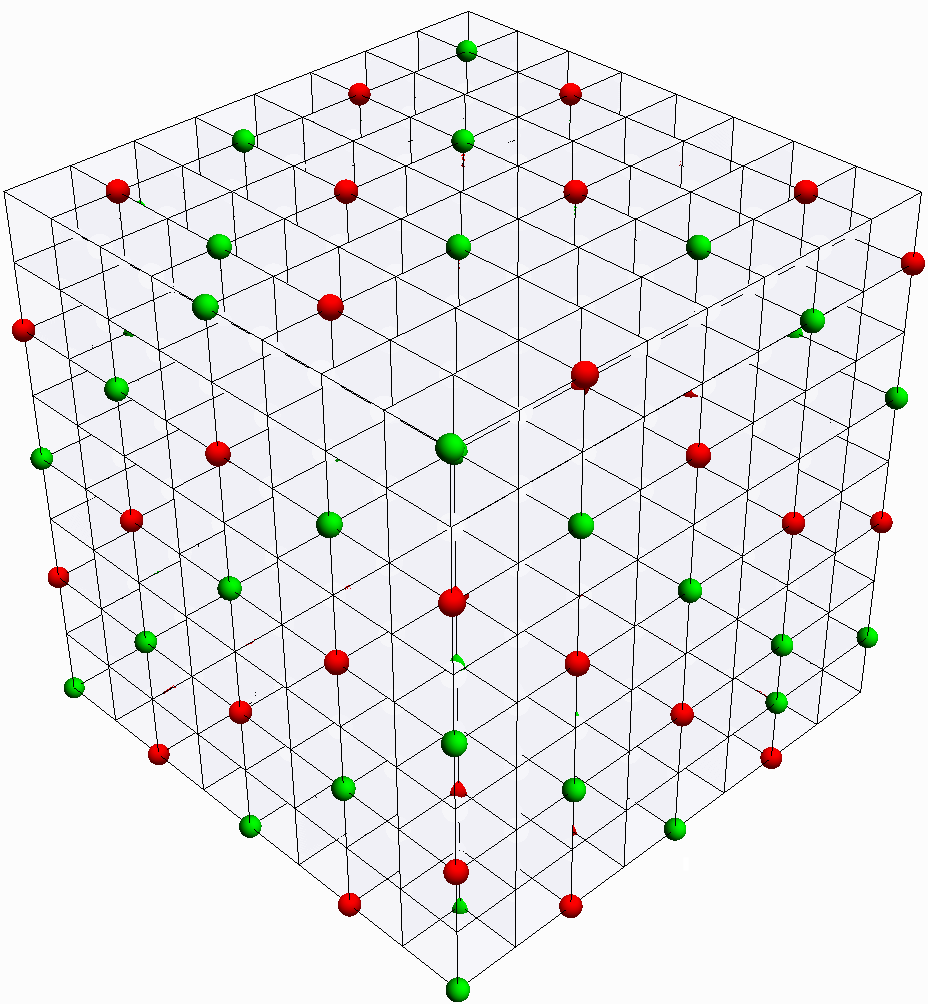}}
    \caption{\label{RealSpaceLattice}(a) A perfect $4Q$ arrangement of hedgehog (red) and antihedgehog (green) singularities in lattice cells that corresponds to the $Q=\pi/2$ state of Fig.\ref{ChiralityPlots}(c). (b) A sample arrangement of singularities on the \emph{dual} cubic lattice, computed with (\ref{TopCharge}) in the equivalent state from a Monte Carlo calculation.}
\end{figure}

To determine the existence of a hedgehog phase, we compute the spin chirality and spin configuration both in real and momentum space. Our main finding is a robust hedgehog lattice phase covering a large portion of the phase diagram. The fast Fourier transform (FFT) of the real-space configurations shows prominent peaks at eight wavevectors in the first Brillouin zone,
\begin{equation}\label{Qpts}
{\bf Q} = Q (n_x \hat{\bf x} + n_y \hat{\bf y} + n_z \hat{\bf z}) \quad,\quad n_x,n_y,n_z \in \lbrace \pm 1 \rbrace \ ,
\end{equation}
whose magnitude given by $Q$ varies as a function of the chiral spin interactions. The FFT peaks are seen in all measured quantities, most notably the chirality components $\chi_{ijk}$ shown in Fig.\ref{ChiralityPlots}, and topological charges (\ref{TopCharge}). This identifies a topological defect lattice in the magnetically ordered phase. The order parameter is periodic, so the state comprises a bipartite lattice of hedgehogs and antihedgehogs organized as a NaCl crystal, i.e. a face-centered cubic lattice. We explicitly visualized this in Fig.\ref{RealSpaceLattice} by computing the topological charges, although the states obtained with Monte Carlo exhibit some amount of disorder in the defect positions. A Hedgehog lattice of this kind is dubbed ``tetrahedral'' $4Q$ in the literature because it can be viewed as a ``superposition'' of four spin spirals aligned with the four apical directions $\pm{\bf Q} \parallel (\pm1,\pm1,\pm1)$ of a regular tetrahedron whose vertices are hedgehogs and the center is an antihedgehog. This hedgehog lattice has been observed experimentally \cite{Fujishiro2019, Nakajima2023} in MnSi$_{1-x}$Ge$_x$ within the composition range $0.3<x<0.6$, and also in SrFeO$_3$ \cite{Tokura2020b}. The latter is a centrosymmetric material, so the Dzyaloshinskii-Moriya interaction $D$ cannot be responsible for the appearance of its hedgehogs. Instead, the higher-order mirror-symmetric chiral interaction $\Phi$ is a candidate agent for hedgehogs in SrFeO$_3$, although it should be noted that we obtain a hedgehog lattice even at $T=0$, while it is experimentally found only at finite temperatures (including in zero magnetic field). A more accurate model is needed to reproduce all details from the experiment.

\begin{figure}[!t]
\centering
    \includegraphics[height=1.7in]{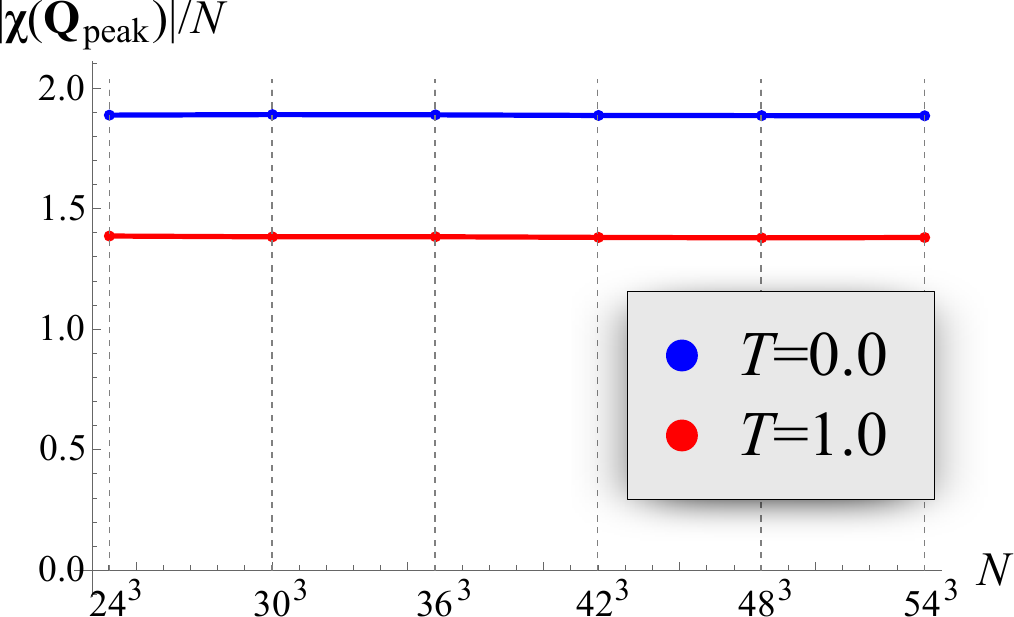}
    \caption{\label{ChiFFT-vs-N}Chirality $\chi$ peak FFT magnitude per lattice site as a function of the system size $N$, shown in the hedgehog lattice at two temperatures, $T=0$ (blue) and $T=1$ (red). The Fourier transforms of the chirality components were computed at the ordering wavevectors ${\bf Q}_{\textrm{peak}}=(\pi/3)\times(\pm1,\pm1,\pm1)$ of the ground-state hedgehog lattice with $J=D=\Phi=1$ and $K=C=0$. The Monte Carlo simulations took about $300$ measurements of the spin configuration in $1500\times N$ iterations, with $23\%$ acceptance of single-spin updates at $T=1$ (simulated annealing was performed at $T=0$).}
\end{figure}

\begin{figure}[!t]
    \includegraphics[width=3in]{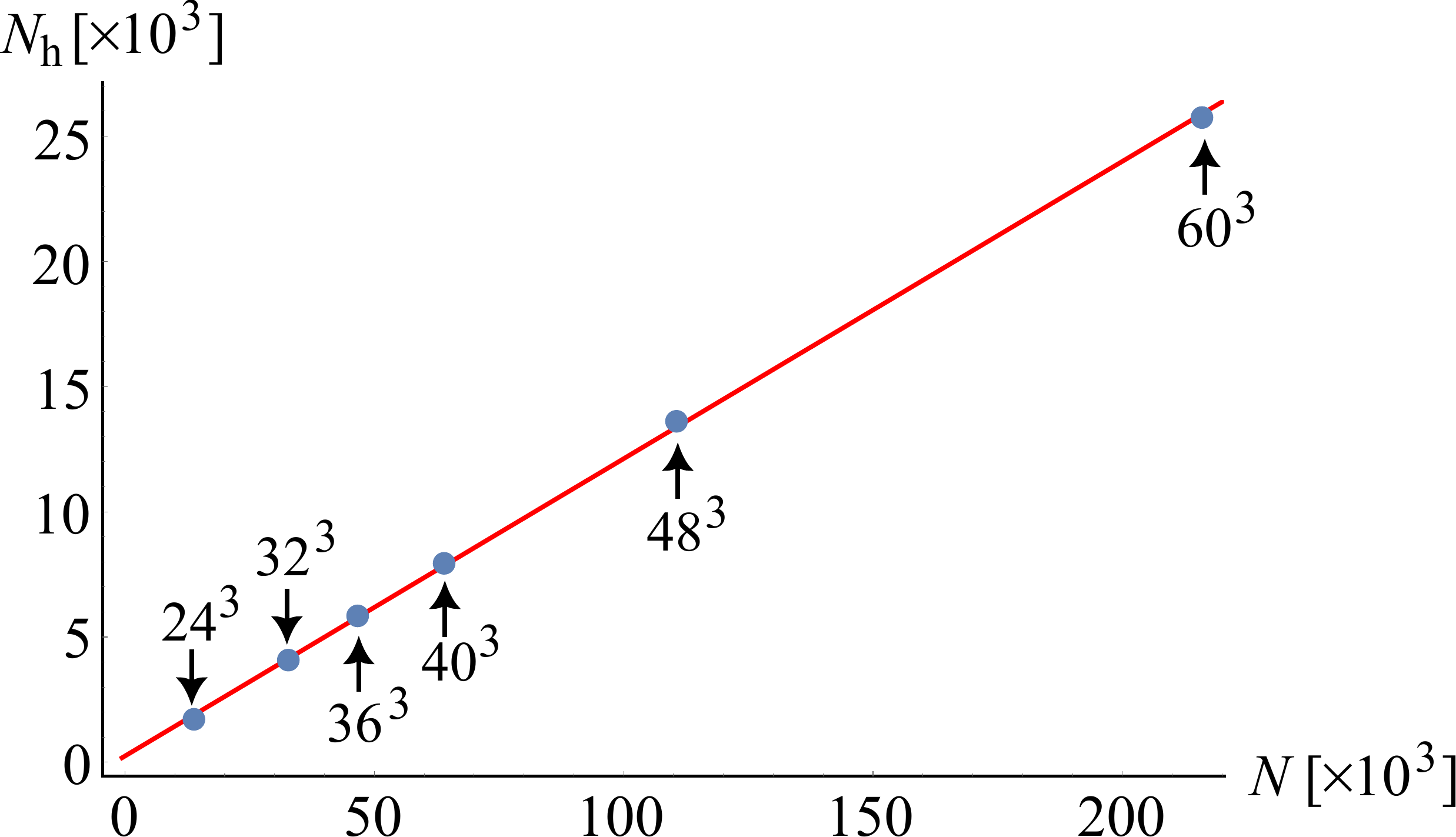}
    \caption{\label{CountN}The number of hedgehogs (and antihedgehogs) in the state of Fig.\ref{ChiralityPlots}(c) as a function of the system size $N$, obtained by simply counting the number (\ref{TopCharge}) of singularities with $\mathcal{N}=+1$ in  the crystal lattice unit-cells. This method slightly overestimates the expected number $N_{\textrm{h}}=N/(2\times 2^3)$ for the given $Q=\pi/2$ metastable hedgehog lattice, due to disorder with random embeddings of extra hedgehog-antihedgehog dipoles.}
\end{figure}

In order to verify the long-range order of the hedgehog lattice, we performed a finite-size scaling. Ideally, one wishes to vary the system size in steps which are compatible with the hedgehog lattice across periodic boundary conditions. The ground-state hedgehog lattice at the parameters $J=D=\Phi=1$ orders at the wavevectors (\ref{Qpts}) with $Q=\pi/3$, so we considered system sizes $N \in \lbrace 24^3, 30^3, 36^3, 42^3, 48^3, 54^3\rbrace$ that can fit this lattice without domain walls. Fig.\ref{ChiFFT-vs-N} shows the peak magnitude of the chirality Fourier transform in the 1st Brillouin zone as a function of the system size $N$. Since this peak magnitude scales in proportion to the system size, the hedgehog lattices obtained with Monte Carlo exhibit long-range order. We find analogous scaling even in metastable states. Fig.\ref{CountN} shows the raw count of the hedgehog + singularities in the lattices ordering at $Q=\pi/2$ when $J=D=\Phi=1$, such as the one depicted in Fig.\ref{ChiralityPlots}(c); despite having a higher energy than the $Q=\pi/3$ state, this hedgehog lattice also exhibits a long-range order under finite-size scaling. Overall, the easy-to-calculate spin chirality alone can provide a reliable characterization of the long-range order in the topological defect lattice.

Note that the defect numbers of Fig.\ref{CountN} were obtained by counting the $+1$ (or alternatively $-1$) singular topological charges in the frozen ground-state configurations at $T=0$, according to (\ref{TopCharge}). Most cubic crystal cells contain zero topological charge, and the plotted fractions of cells in Fig.\ref{CountN} contain the $\pm1$ charge. The cells with other integer values of topological charge are rare and their number does not scale with the system size. The total topological charge in the system is also pinned to zero by the periodic boundary conditions in the simulations.

\begin{figure}[!]
\centering
    \subfigure[{}]{
      \includegraphics[width=3.1in]{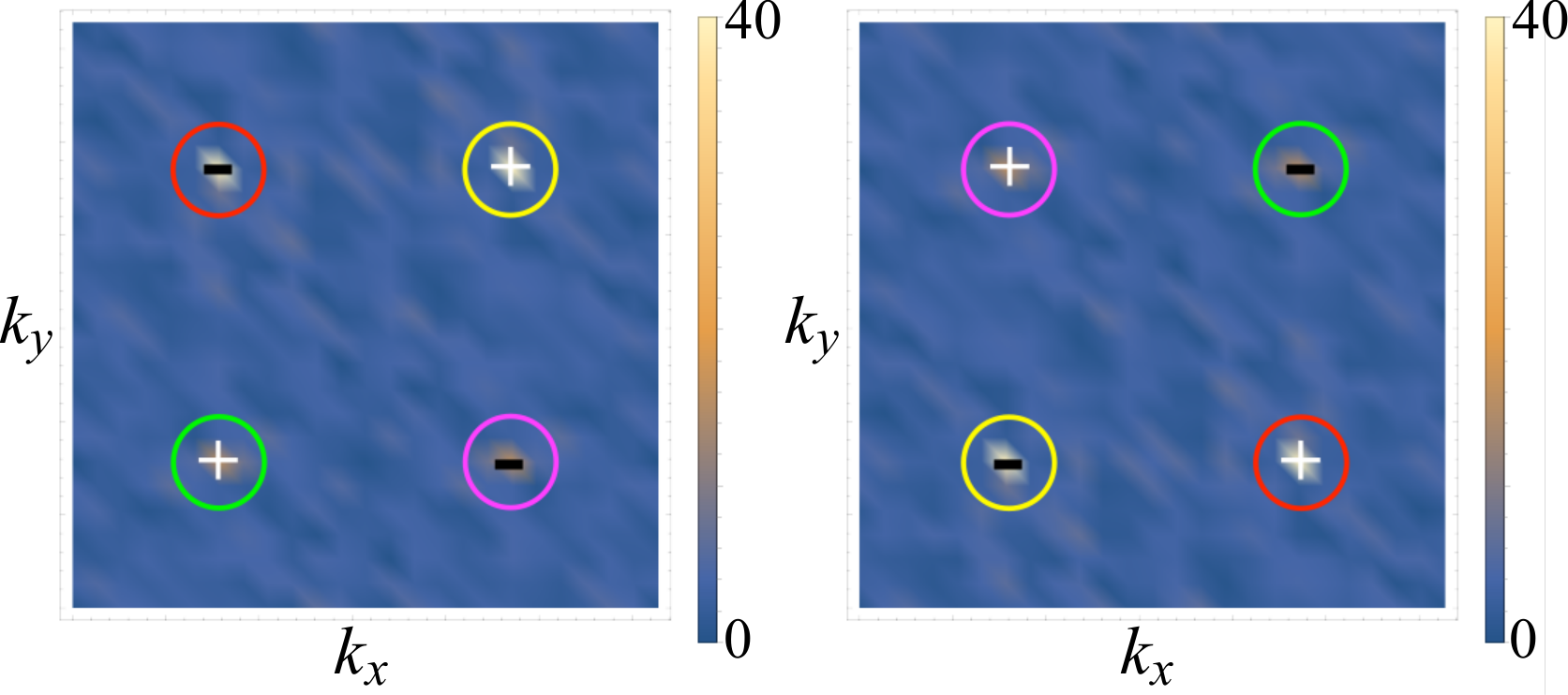}} 
    \subfigure[{}]{
       \includegraphics[width=3.1in]{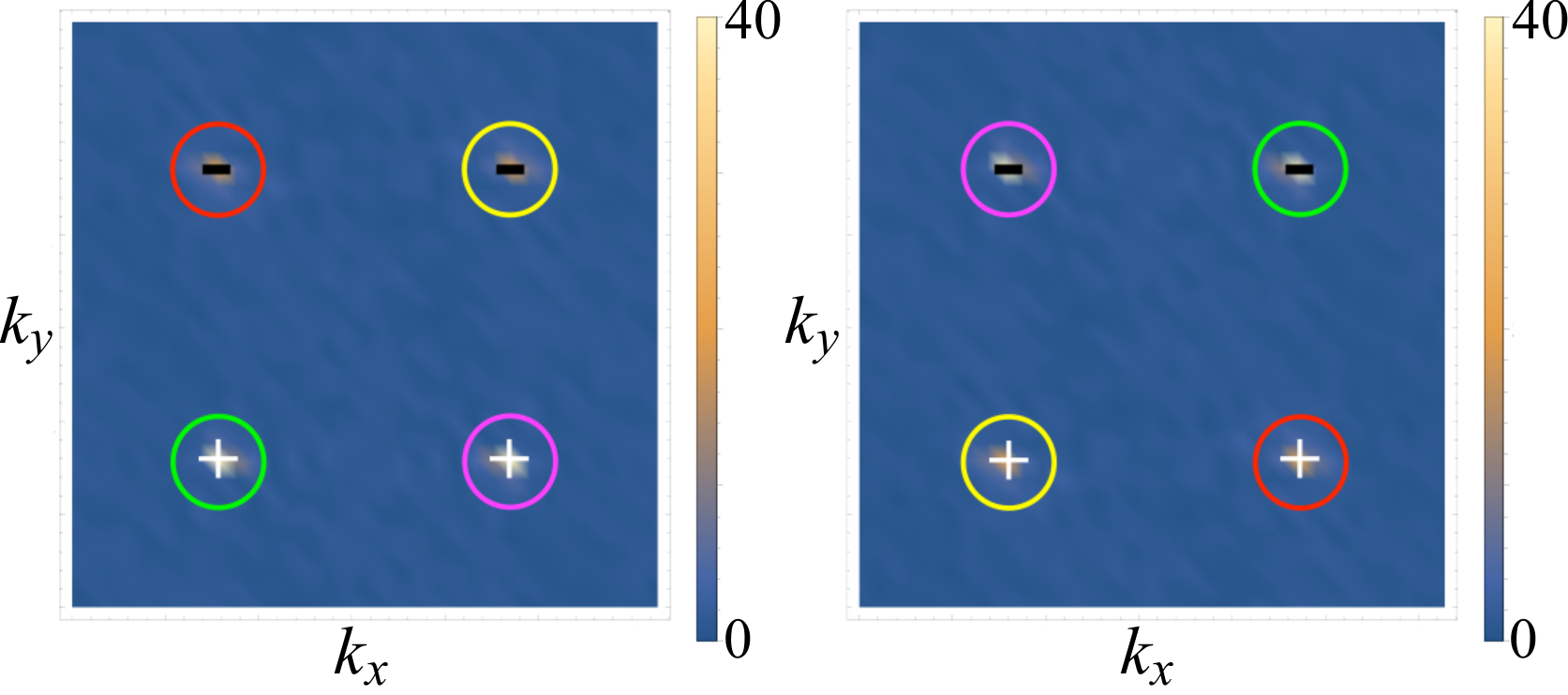}}
    \caption{\label{ChiralPhases}Typical relative phases of the complex-valued FFT peaks in a single chirality pseudovector component, for the state shown in Fig.\ref{ChiralityPlots}(c). The two depicted patterns, observed in calculations with system sizes (a) $N=24^3$ and (b) $N=32^3$, are randomly selected by the precise manner in which the translational symmetry is broken by the hedgehog lattice. The left and right columns are obtained at $k_z=+\pi/2$ and $k_z=-\pi/2$ respectively. The circled signs denote sign changes of the phase (the same-colored circles in each row have opposite signs because the real-space chirality is real).}
\end{figure}

Fig.\ref{ChiralPhases} shows relative phases of the FFT peaks in a single chirality component at wavevectors (\ref{Qpts}) for the state depicted in Fig.\ref{ChiralityPlots}(c). The patterns are consistent with the $4Q$ structure. All observed variations correspond only to the translations of the real-space order parameter along the principal lattice directions, and not changes in the point-group symmetries of the hedgehog lattice.

\begin{figure*}[!]
\subfigure[{}]{\includegraphics[width=1.7in]{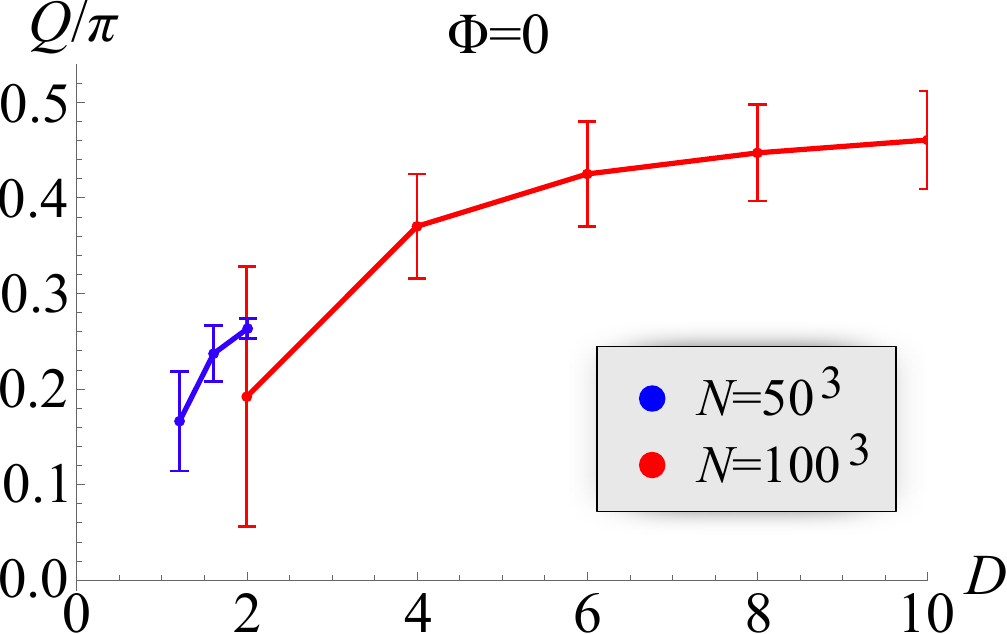}}\hspace{0.05in}
\subfigure[{}]{\raisebox{-0.12in}{\includegraphics[width=1.7in]{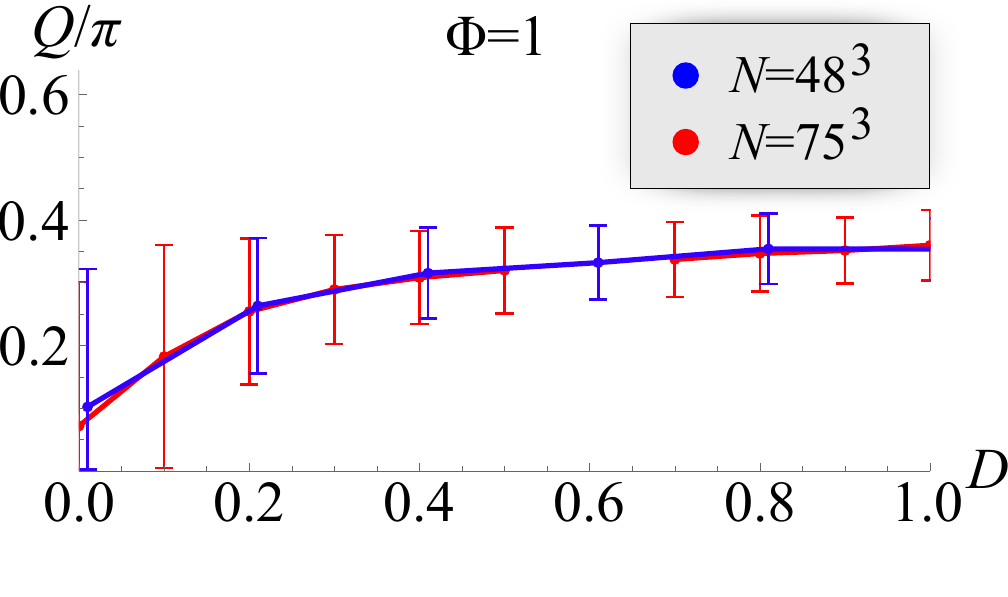}}}\hspace{0.05in}
\subfigure[{}]{\includegraphics[width=1.7in]{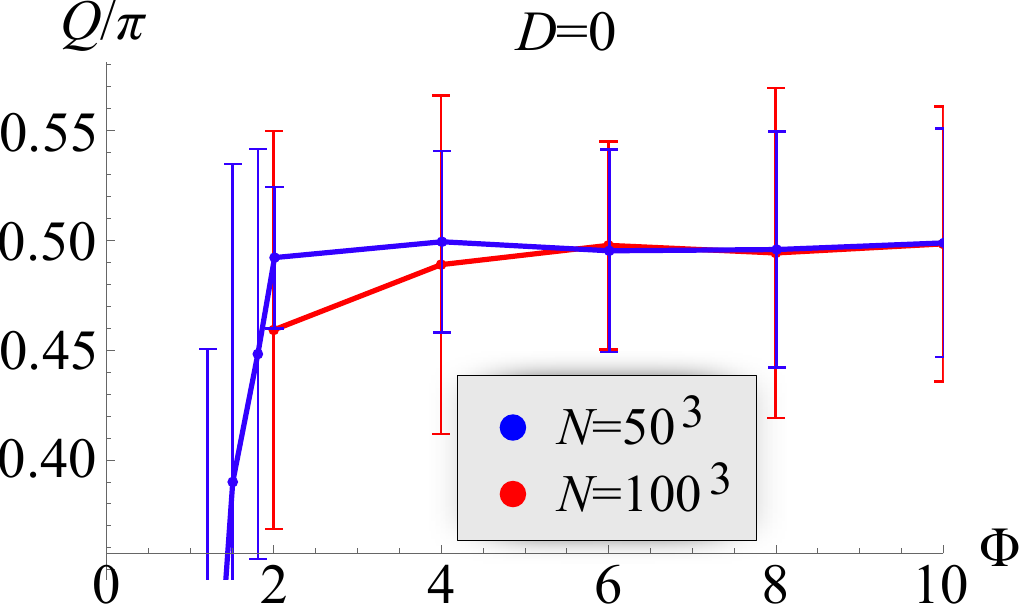}}\hspace{0.05in}
\subfigure[{}]{\includegraphics[width=1.7in]{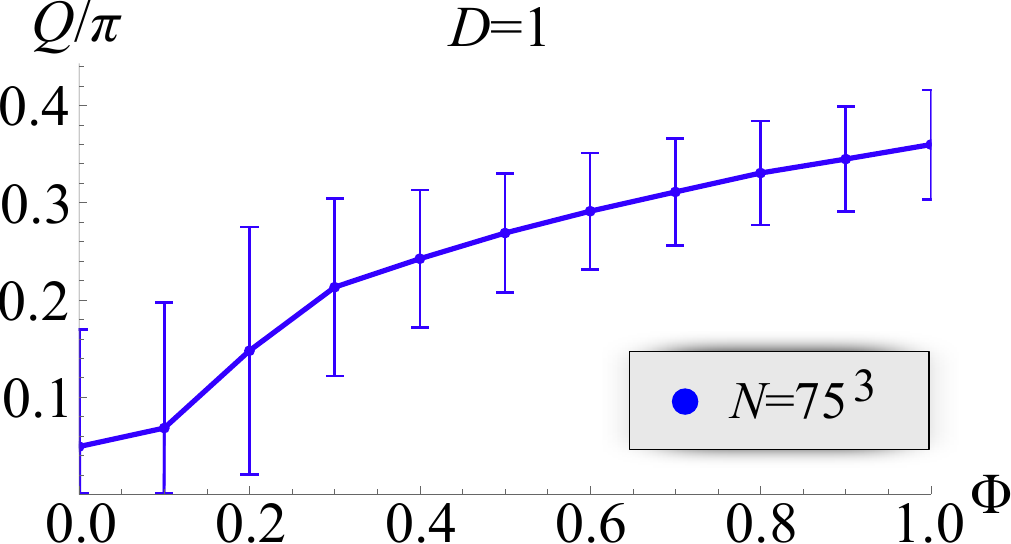}}
\subfigure[{}]{\includegraphics[width=1.7in]{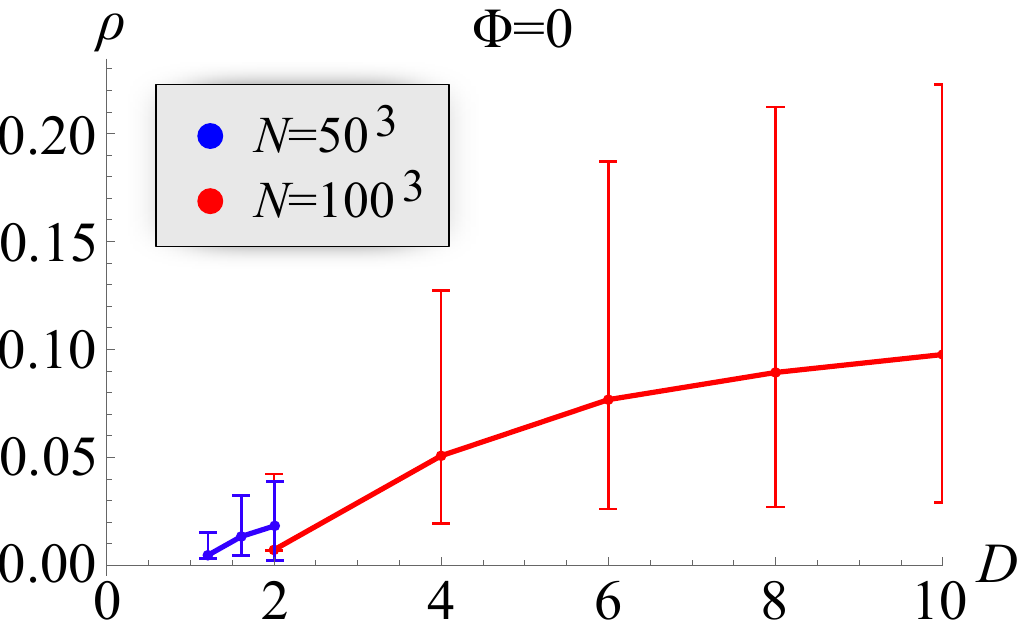}}\hspace{0.05in}
\subfigure[{}]{\includegraphics[width=1.7in]{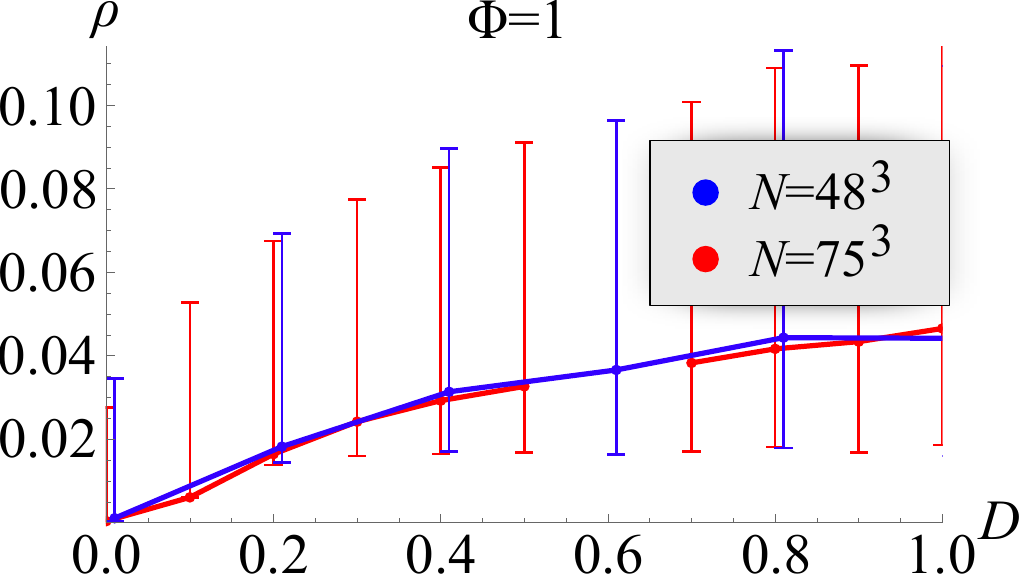}}\hspace{0.05in}
\subfigure[{}]{\includegraphics[width=1.7in]{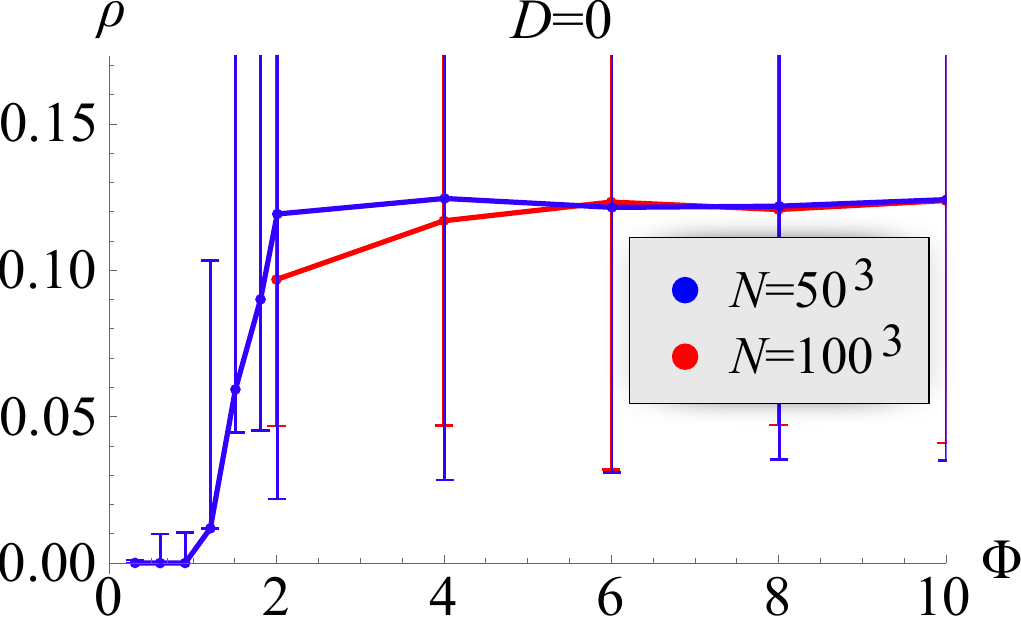}}\hspace{0.05in}
\subfigure[{}]{\includegraphics[width=1.7in]{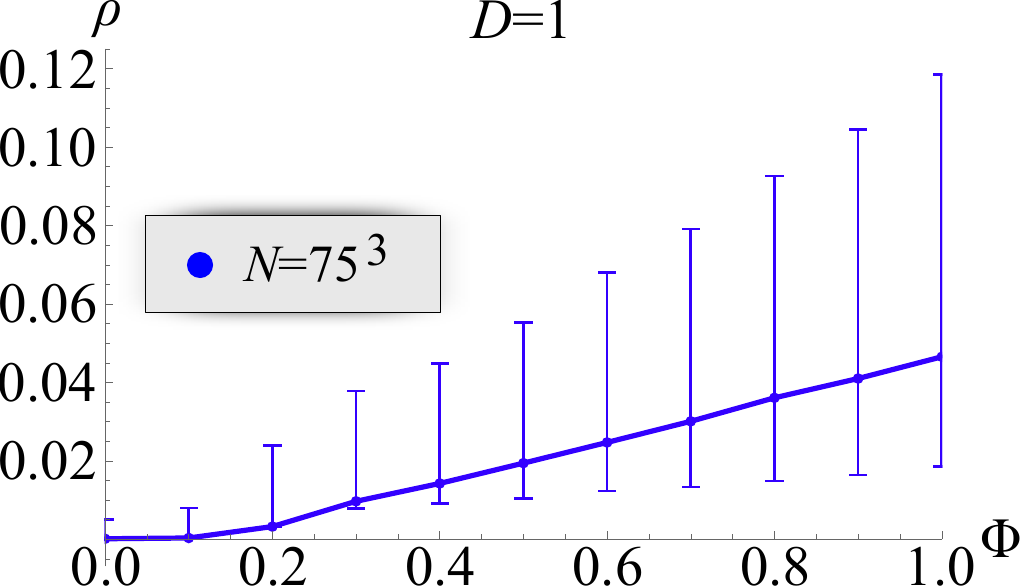}}
\caption{\label{Hlat-density}The dependence of the hedgehog lattice ordering wavevector ${\bf Q}=Q(\pm1,\pm1,\pm1)$ and the corresponding $\mathcal{N}=+1$ hedgehog density $\rho$ on the chiral interactions $D$ (Dzyaloshinskii-Moriya) and $\Phi$ (4-spin). The ordering wavevector magnitude $Q$ and its uncertainty $\delta q$ (represented by error bars) were obtained from the Lorentzian fit $A+B/\lbrack({\bf k}-{\bf Q})^2+\delta q^2\rbrack$ to the magnitude of the chirality Fourier transform $|\boldsymbol{\chi}({\bf k})|$. Note that $\delta q$ is a reflection of metastability involving frozen disorder and domain walls, i.e. the limited time spent on annealing and measurements. Domain walls might have been artificially created in many runs by not matching the system size to the (initially unknown) hedgehog lattice period; this problem was eliminated in certain selected simulations, see Fig.\ref{SpinChirality}(b,c). The error bars for the hedgehog density $\rho\propto Q^3$ are determined by $\rho(Q\pm\delta q) = \rho(Q)\pm\delta\rho$. Data was collected from various simulations with indicated system sizes $N$ and $J=1$, $K=C=0$, mostly at $T=0$ after simulated annealing.}
\end{figure*}

\begin{figure}[!]
\centering
    \subfigure[{}]{\includegraphics[width=2.2in]{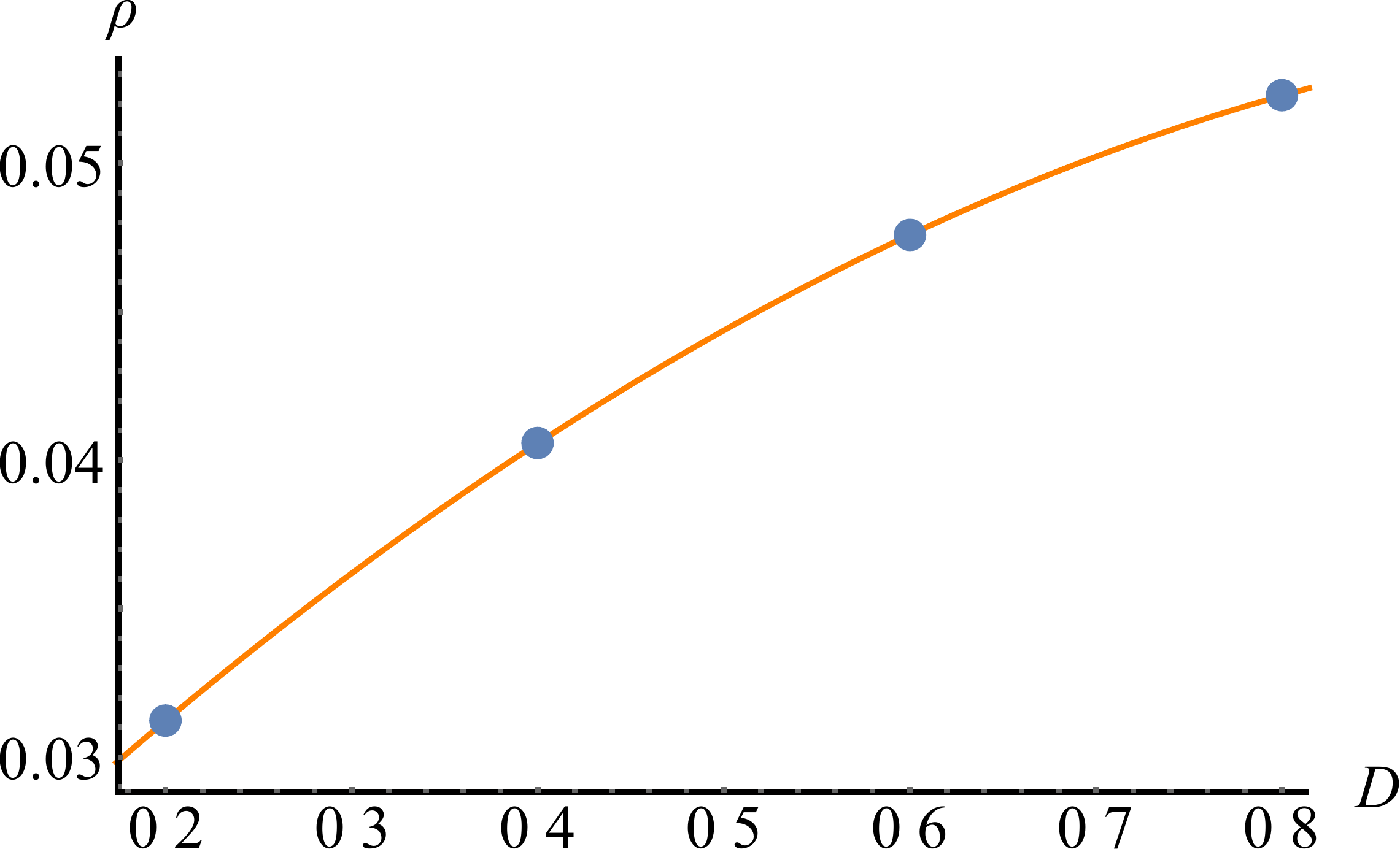}}
    \subfigure[{}]{\includegraphics[width=2.2in]{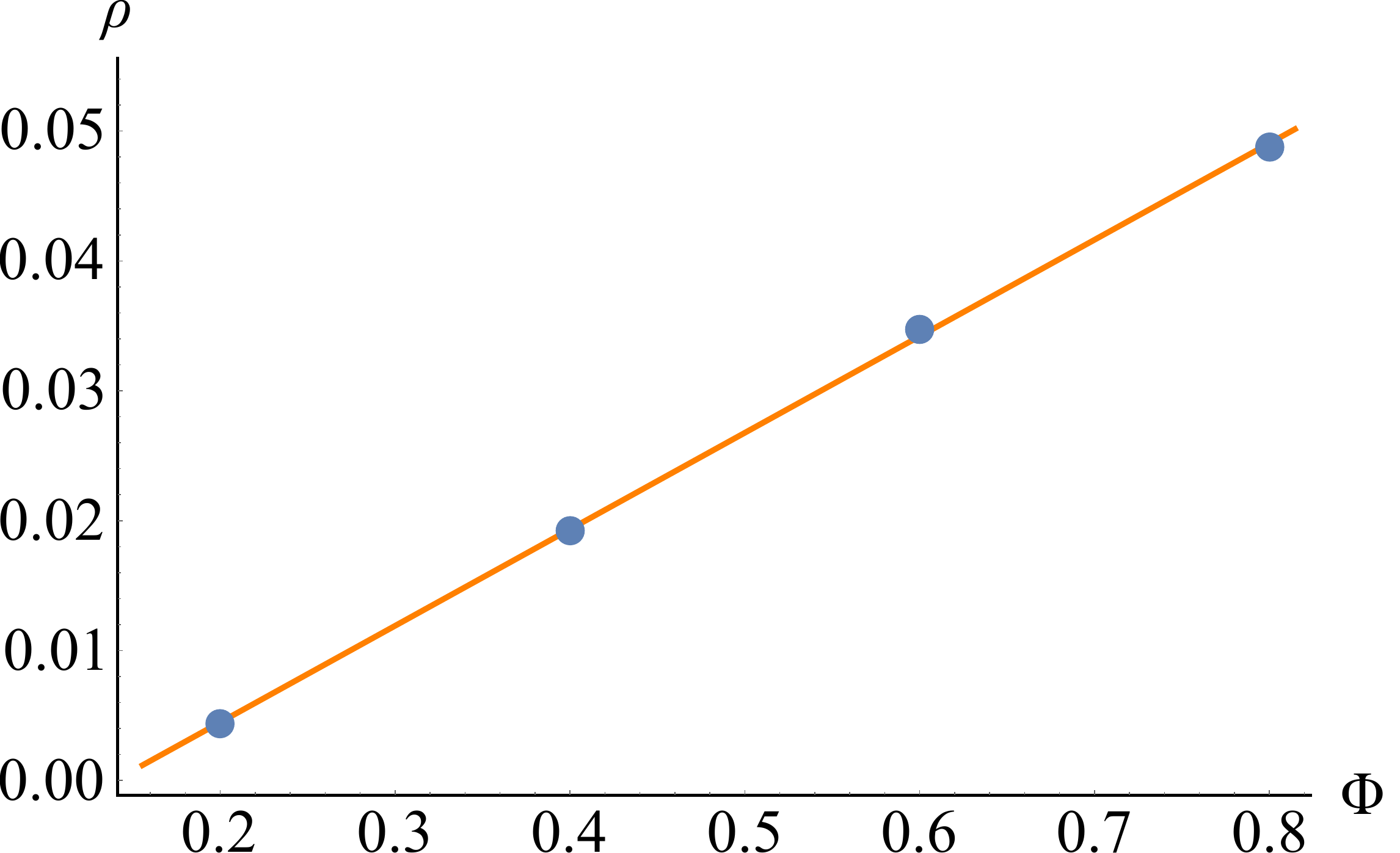}}
    \caption{\label{RhovsDPhi}The density $\rho$ of hedgehogs (a) as a function of the isotropic DM interaction $D$ at $\Phi=1$ and (b) as a function of the 4-spin chiral interaction $\Phi$ at $D=1$. Both plots are generated by counting the raw number of $\mathcal{N}=+1$ hedgehog singularities (\ref{TopCharge}) in the crystal lattice unit-cells, using simulations with $J=1$, $K=C=0$, temperature $T=0$ (after simulated annealing) and system size $N=75^3$. This method only slightly overestimates but systematically follows the densities extracted from the Lorentzian-fit peaks in Fig.\ref{Hlat-density}, indicating that the large error bars in Fig.\ref{Hlat-density} significantly exaggerate the uncertainty of the ordering wavevector.}
\end{figure}

The density of hedgehogs and antihedgehogs is controlled by the isotropic chiral interactions (\ref{DMInteraction}) and (\ref{ChiralInteraction}). Both are independently able to stabilize hedgehog lattices, but seemingly only if they have a sufficient finite strength. Fig.\ref{Hlat-density} shows how the hedgehog lattice ordering wavevector and the corresponding hedgehog density depend on the Dzyaloshinskii-Moriya $D$ and 4-spin $\Phi$ chiral interactions. This data was collected from numerous Monte Carlo runs and extracted by Lorentzian fits to the peaks of the chirality Fourier transform magnitude. The error bars, given by the Lorentzian width, appear to be large due to the significant amount of disorder in the hedgehog lattices, especially in larger systems and at small $D,\Phi$ values. However, the Lorentzian peak position, which gives the ordering wavevector and hedgehog density, is more reliable. Fig.\ref{RhovsDPhi} depicts a different estimate of the hedgehog density, obtained by simply counting the hedgehogs as $\mathcal{N}=+1$ spin configurations on the cubic unit-cells of the crystal. The hedgehog densities estimated using these two different methods agree.

We can make several physical observations from these results. First, the Dzyaloshinskii-Moriya interaction $D$ produces a non-linear growth of the defect density with the interaction strength. In contrast, the isotropic multiple-spin chiral coupling $H_{\textrm{C}}$ is the real ``chemical potential'' for the point defects, giving rise to a linear relationship between the defect density and interaction strength. This intuition also follows from the field theory considerations, where the topological charge of hedgehogs is captured by a rank-2 current compatible with a rank-2 gauge field of the multiple-spin chiral interaction. Physically, the $H_{\textrm{C}}$ term wants to align the spatial orientation of the spin chirality (\ref{SpinChirality}) with the local magnetization background. This is possible only for $\pm1$ singularities. The $+1$ hedgehogs and $-1$ antihedgehogs are equivalent in this respect in all odd spatial dimensions $d$, but both must be nucleated and evenly intermixed in order to maximize the number of singularities without producing large spatial regions whose total topological charge exceeds $\pm1$. Ultimately, the chiral interactions act as a chemical potential for hedgehogs, which explains the linear dependence of the hedgehog density on $\Phi$.

The intricate landscape of hedgehog lattices with different ``lattice constants'' opens a possibility of 1st order phase transitions between different lattice configurations. At large hedgehog densities we explore, it is expected that all hedgehog lattices are commensurate with the microscopic crystal lattice, so that their spacing cannot continuously evolve in response to the gradual changes of $D$ and $\Phi$. Unfortunately, mapping the low-temperature phase diagram of these states is currently a formidable task, left for future work. The plots of $\rho(D,\Phi)$ naively suggest that the hedgehog density $\rho$ vanishes when both $D$ and $\Phi$ are sufficiently small but still finite. While this could be a physical effect, our present numerics cannot rule out the possibility that stable hedgehog lattices of arbitrarily low density exist with arbitrarily small $D, \Phi$. In order to verify this possibility, it is necessary to carefully average-out the positional disorder of hedgehogs in their lattice, i.e. significantly reduce the error bars in Fig.\ref{Hlat-density}.

Since the numerical calculations are time-consuming, we have sampled only a limited portion of the full parameter space and found in it only one type of an ordered phase with a hedgehog lattice. Other structures could be stable with other model parameter choices. Some calculations have produced a few prominent metastable hedgehog lattices with more than eight FFT peaks in the first Brillouin zone, but only in small systems (i.e. the additional peaks do not scale with the system size).

\subsection{Finite-temperature phase transition}\label{secPhTrans}

\begin{figure}[!t]
\centering
  \includegraphics[width=3.1in]{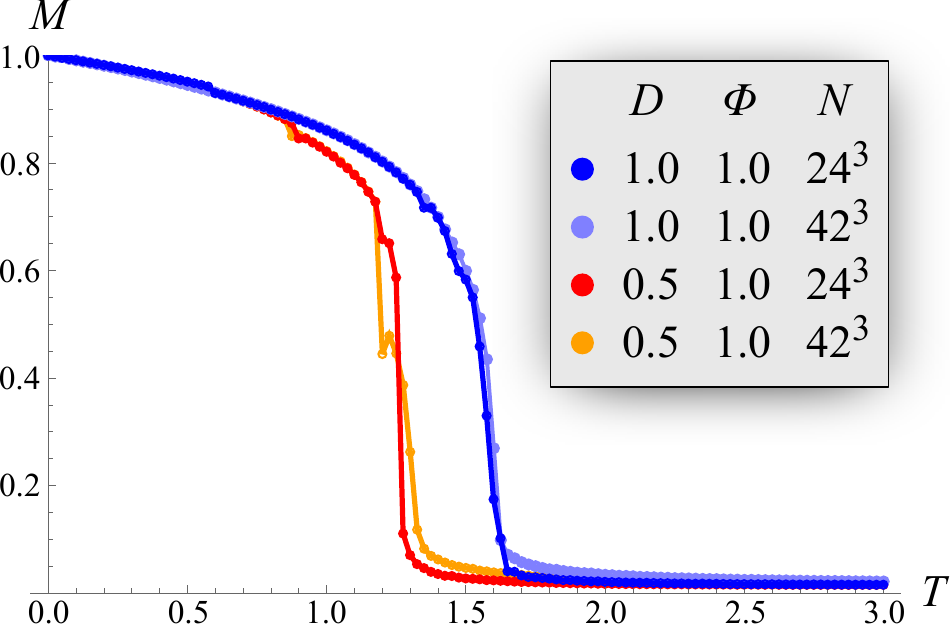}
  \caption{\label{Magnetization}Thermally and lattice-averaged local magnetization magnitude (\ref{M_vs_T} as a function of temperature for two combinations of chiral couplings and two system sizes ($J=1$, $K=C=0$). These Monte Carlo simulations took $2000$ measurements over $5\times 10^8$ iterations in equilibrium at every temperature point, each starting from the microstate achieved at the previous higher temperature point in a gradual temperature down-sweep with given $D,\Phi$.}
\end{figure}

Monte Carlo simulations indicate that a single phase transition occurs at a critical temperature $T_c$, as shown in Fig.\ref{Magnetization}. The high-temperature phase looks like a conventional paramagnetic state with a vanishing local magnetization. This transition is unambiguously of the second order if no hedgehog lattice exists in the low-temperature magnetic state, i.e. if $D=\Phi=0$. A second-order transition is indeed known to occur in the ordinary Heisenberg model on the cubic lattice. However, if the low-temperature phase hosts a hedgehog lattice, we sometimes observe a magnetization jump at $T=T_c$, which is consistent with a first-order transition.

Fig.\ref{Magnetization} shows the temperature dependence of the magnetization ``order parameter'' defined as
\begin{equation}\label{M_vs_T}
M = \frac{1}{N}\sum_{i}|\langle{\bf S}_{i}\rangle|^{2} \ ,
\end{equation}
where $\langle\cdots\rangle$ indicates thermal average. Many samples of the spins ${\bf S}_i$ are collected during the Monte Carlo simulation at fixed $T$, and their thermal fluctuations are first averaged out independently on every lattice site $i$. Generally, $\langle{\bf S}_{i}\rangle\to0$ in disordered phases when the ${\bf S}_i$ samples are collected over a sufficiently long amount of time $\tau$. Then, the magnitude of these thermally-averaged spins is averaged out over all lattice sites to obtain the ``order parameter'' $M(T)$. This quantity is useful because it reveals the presence of time-reversal-breaking magnetic order regardless of the precise spin texture it realizes in space. In fact, we do not know what particular spin texture to anticipate in the ground state; much microscopic freedom for local spin orientations remains even after knowing the structure of the hedgehog lattice. The price to pay is that $M(T)$ can develop a non-analytic form, signalling a phase transition, only under a simultaneous scaling of the system size $N\to\infty$ and sampling time $\tau\to\infty$. The importance of $\tau$ is evident in Fig.\ref{Magnetization}: the transitions look less sharp in larger systems when the sampling time $\tau$, i.e. the number of Monte Carlo iterations, is not scaled-up together with the system size $N$.

\begin{figure}[!t]
\centering
  \subfigure[{}]{\includegraphics[width=3.1in]{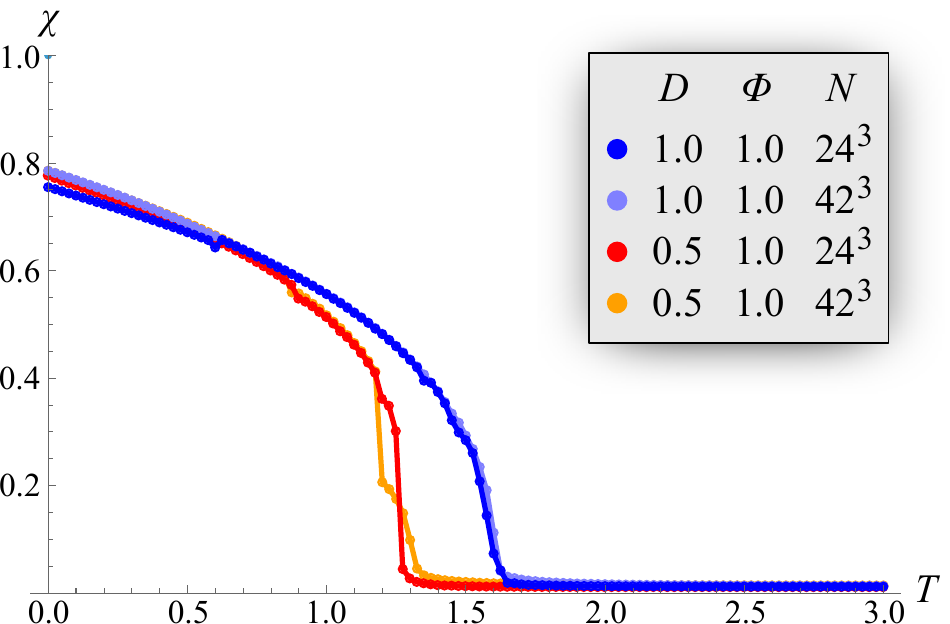}}
  \subfigure[{}]{\includegraphics[width=3.1in]{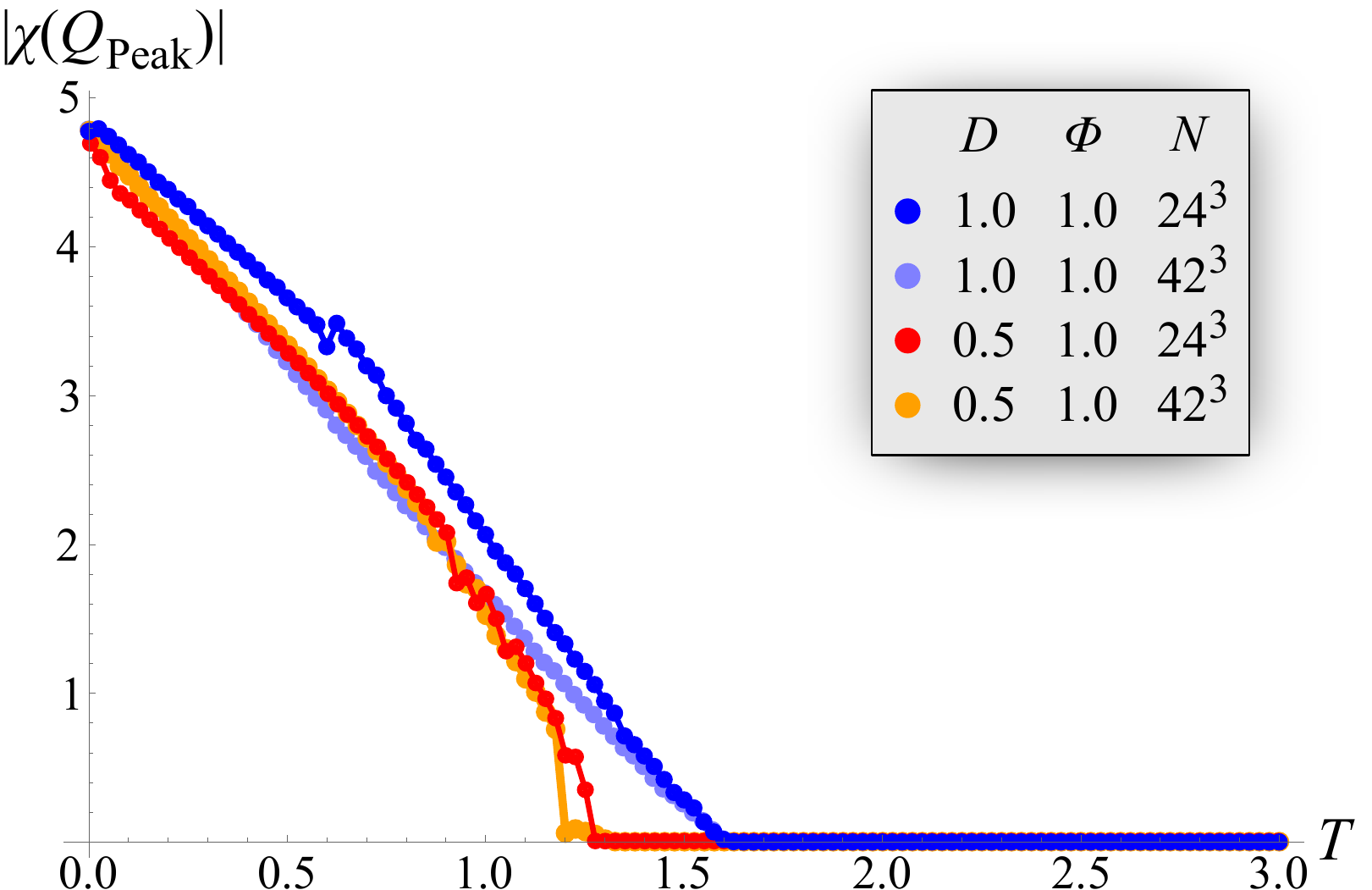}}
  \caption{\label{Chirality}Temperature dependence of the (a) thermally and lattice-averaged local chirality magnitude (\ref{Chi_vs_T}, and (b) average magnitude of the eight chirality FFT peaks at the ordering wavevectors (\ref{Qpts}) with $Q=\pi/3$. The Monte Carlo data set is the same as in Fig.\ref{Magnetization}.}
\end{figure}

\begin{figure}
  \subfigure[{}]{\includegraphics[width=3.1in]{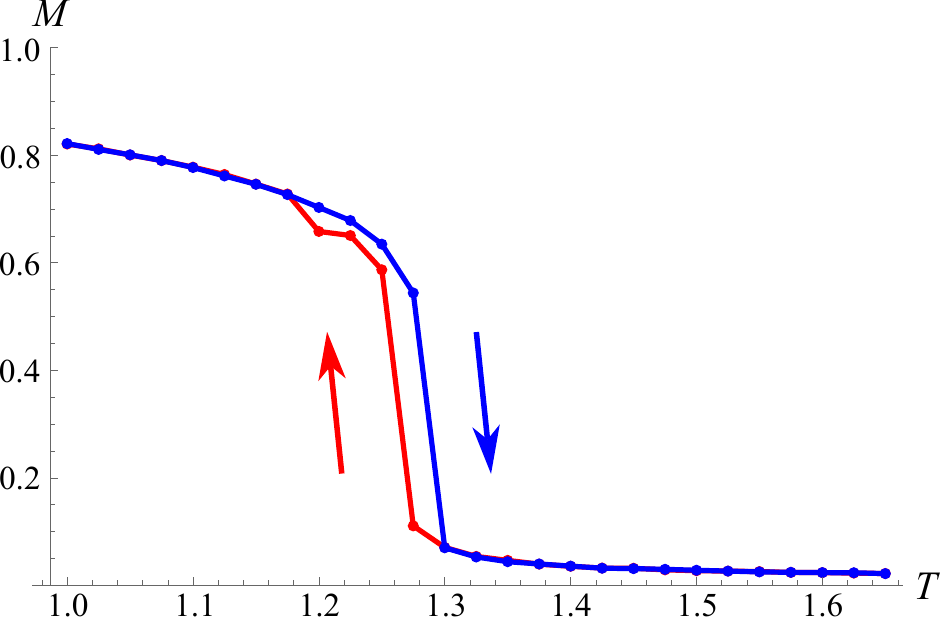}}
  \subfigure[{}]{\includegraphics[width=3.1in]{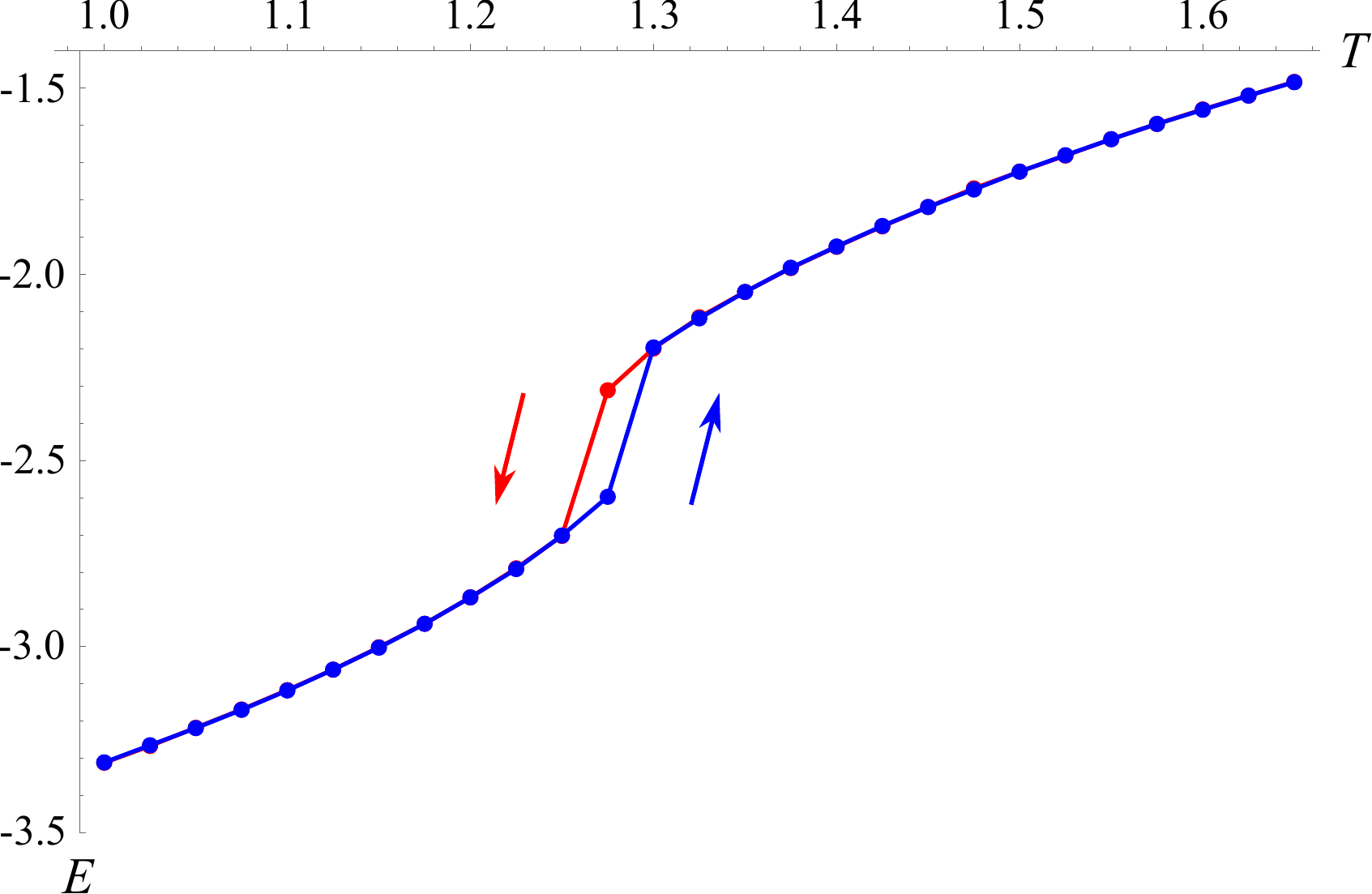}}
  \caption{\label{Hysteresis}Hysteresis in the vicinity of the phase transition at $D=0.5$, $\Phi=1$ ($J=1$, $K=C=0$) is visible in the (a) magnetization (\ref{M_vs_T}) and (b) internal energy (per lattice site) when the temperature down-sweep and up-sweep are compared. The temperature down-sweep is the $N=24^3$ data from the Fig.\ref{Magnetization}. The final $T=0$ microstate of the down-sweep was used to seed the beginning of the up-sweep, and the same Monte Carlo simulation parameters were applied in all runs.}
\end{figure}

Our results suggest that first and second order phase transitions might occur at a critical temperature in different regions of the phase diagram. The transition appears to be of first order when $\Phi=1$, $D=0.5$. This follows from consistent discontinuities of the magnetization (Fig.\ref{Magnetization}) and chirality (Fig.\ref{Chirality})
\begin{equation}\label{Chi_vs_T}
\chi = \frac{1}{N}\sum_{i}|\langle{\boldsymbol{\chi}}_{i}\rangle|^{2}
\end{equation}
order parameters, as well as internal energy $E$ (Fig.\ref{Hysteresis}), at the critical temperature $T_c\approx 1.275$. The discontinuities are observed in repeated temperature down/up sweeps, with multiple system sizes $N\in\lbrace 18^3, 24^3, 30^3, 42^3\rbrace$ and with different sampling times $\tau$. The data from these runs collapses to the same curves with a negligible dependence on $N$ and $\tau$ in the probed range. Furthermore, a hysteresis shown in Fig.\ref{Hysteresis} appears when the temperature down-sweeps and up-sweeps are compared.

The observed discontinuities and hysteresis are strong indicators of a first order phase transition. The Helmholtz free energy $F=E-TS$ is minimized in equilibrium and hence varies continuously across $T_{c}$. In a first order transition, the free energy $F\lbrack\langle{\bf S}_{i}\rangle\rbrack$ as a functional of the spin configuration $\langle {\bf S}_{i}\rangle$ has degenerate local minimums at $\langle{\bf S}_{i}\rangle\neq0$ and $\langle{\bf S}_{i}\rangle\equiv0$ separated by a barrier. Then, a sudden switch of the spin configuration from one minimum to the other makes the entropy $S(T)$ discontinuous at $T_{c}$, and this implies a discontinuity in the internal energy $E(T)$ directly measured with Monte Carlo, i.e. a latent heat. Note that $E(T)$ would not be discontinuous if the classical first order phase transition occurred at $T=0$, since $F=E$ would then pass on the degeneracy to $E\lbrack\langle{\bf S}_{i}\rangle\rbrack$. Therefore, the internal energy jump shown in Fig.\ref{Hysteresis}, accompanied by an order parameter jump, reveals the existence of degenerate competing macrostates at $T_c$. The resulting bimodal internal energy is also implicitly evident via the hysteresis in Fig.\ref{Hysteresis}(b) at $T\approx T_c$. Note that the discontinuities are seen in very gradual temperature sweeps, using the Metropolis-Hastings algorithm that updates one spin at a time. If a single high-temperature free energy minimum had gradually evolved into a ``Mexican hat'' shape across $T_c$, then the continuously-developing order at the ``Mexican hat'' minimum would have been far more likely to discover with Monte Carlo than a competing metastable ordered state across a free energy barrier (the number of microstates under the barrier is an exponential function of the system size).

Of course, phase transitions cannot occur in finite systems, so it is important to understand the scaling of observables. We have two scaling parameters, the system size $N$ and the sampling time $\tau$. The limit $\tau\to\infty$ corresponds to the ideal statistical averaging of measured quantities, which ensures ergodicity. In this limit, $E(T)$ is analytic at every finite $N$, but develops some singularity at a phase transition when $N\to\infty$. Unlike second order phase transitions, which are characterized by universality due to a divergent correlation length $l\to\infty$, first order transitions manifest correlations of finite range $l<\infty$ that become undisturbed by the system boundary when $N$ becomes large enough, $N\gg l^3$. The slope of $E(T)$ at $T=T_c$ is expected to grow in proportion to $N$ \cite{Kosterlitz1991}. This behavior can be also expected at finite $\tau$ as long $\tau>\tau_0$ is larger than ``ergodicity time'' $\tau_0$ needed to explore all microstates in a Monte Carlo run. Since every macrostate corresponds to an exponential number $e^{bN}$ of microstates, we suspect that $\tau_0$ grows exponentially with the system size $N$. The scaling with $N\to\infty$ will become corrupted if $\tau$ is kept fixed because $\tau$ becomes smaller than $\tau_0(N)$ at some point. In our results, we have presumably achieved the $\tau>\tau_0$ regime with sufficiently small lattices $N\in\lbrace 18^3, 24^3, 30^3\rbrace$ and then observed the discontinuous transition that hardly depends on $N$ because $l<N^{1/3}$. Our attempts to push larger system sizes, $N\ge 36$ have not been equally successful, probably because our practical limitations on $\tau$ have then placed us in the $\tau<\tau_0$ regime; this is presented in the Appendix.

Another physical observation provides further support for the first order transition. The ordered phase is a hedgehog lattice. In continuum space, a hedgehog at position ${\bf R}$ is a singular point of the chirality $\boldsymbol{\chi}({\bf r})\sim\pm({\bf r}-{\bf R})/|{\bf r}-{\bf R}|^{3}$. Hedgehog singularities are regularized on the lattice, but the ensuing spatial variations of the spins are still rapid near the hedgehog cores and relatively smooth only in the further regions between the cores. When temperature increases, the geometrically frustrated spins in the cores are most affected; their thermal fluctuations efficiently average-down the nearly-singular chirality of the cores and lead to a quick ``linear'' decrease of the peak chirality FFT at the ordering wavevector ${\bf Q}_{\textrm{Peak}}$, shown in Fig.\ref{Chirality}(b). Since the chirality identifies hedgehogs, this is physically equivalent to thermally-induced positional fluctuations of the hedgehog cores that chip away from the ordering at ${\bf Q}_{\textrm{Peak}}$. At he same time, the local magnetization magnitude (\ref{M_vs_T}) lingers near its saturated value of $1$ throughout most of the temperature range in which the hedgehog lattice is stable, as shown in Fig.\ref{Magnetization}. Importantly, it is still finite at $T=T_c$. This means that the spins in the further regions between the cores, where chirality is naturally close to zero, manage to resist thermal fluctuations relatively well. It is, therefore, the spins inside the core regions that take the toll of thermal fluctuations, but this is identified with the positional jittering of hedgehogs. In this emerging picture, a phase transition ultimately occurs by hedgehog lattice melting, and melting transitions are generally of first order.

\begin{figure}[!t]
\centering
  \subfigure[{}]{\includegraphics[width=3.5in]{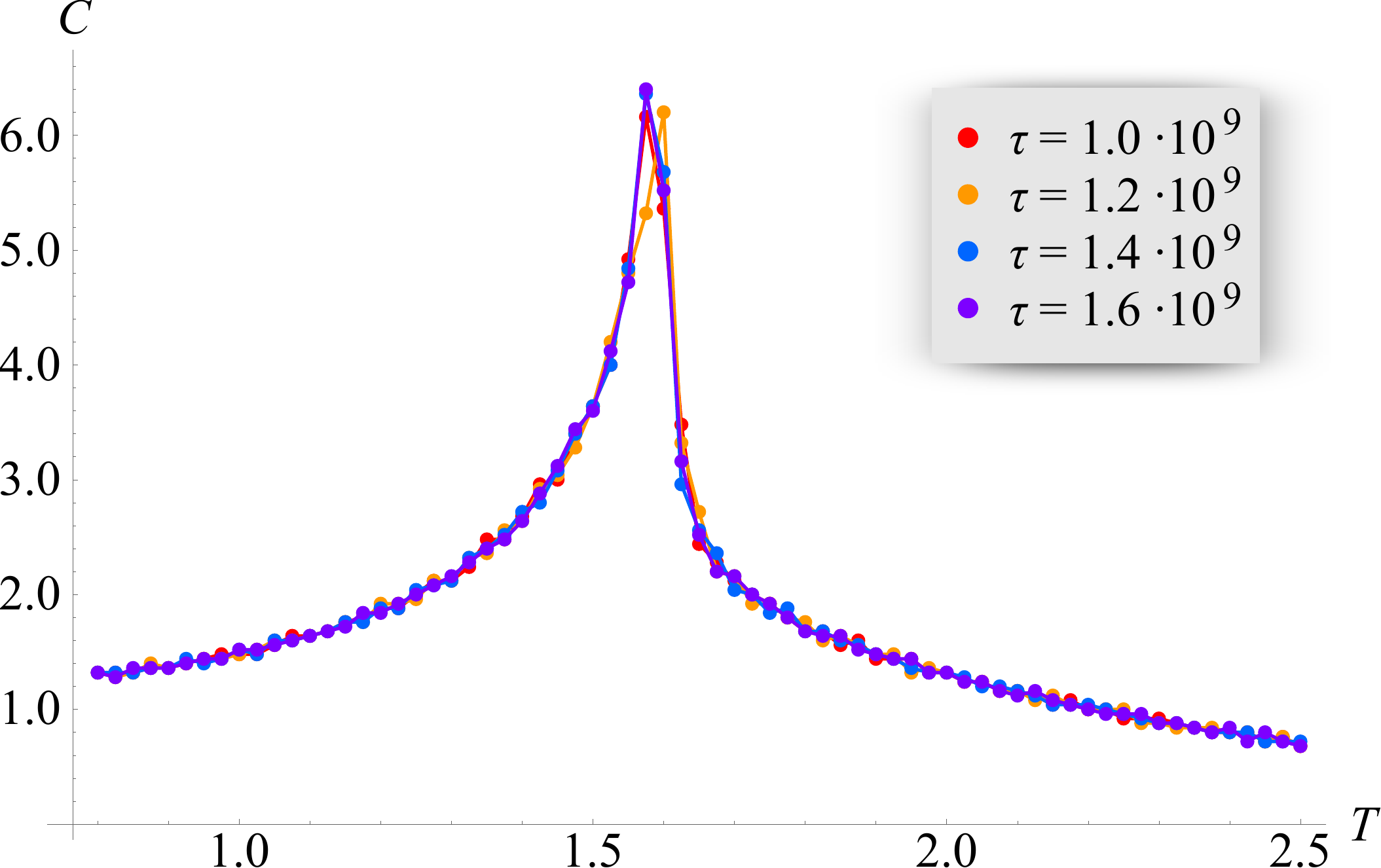}}
  \subfigure[{}]{\includegraphics[width=3.3in]{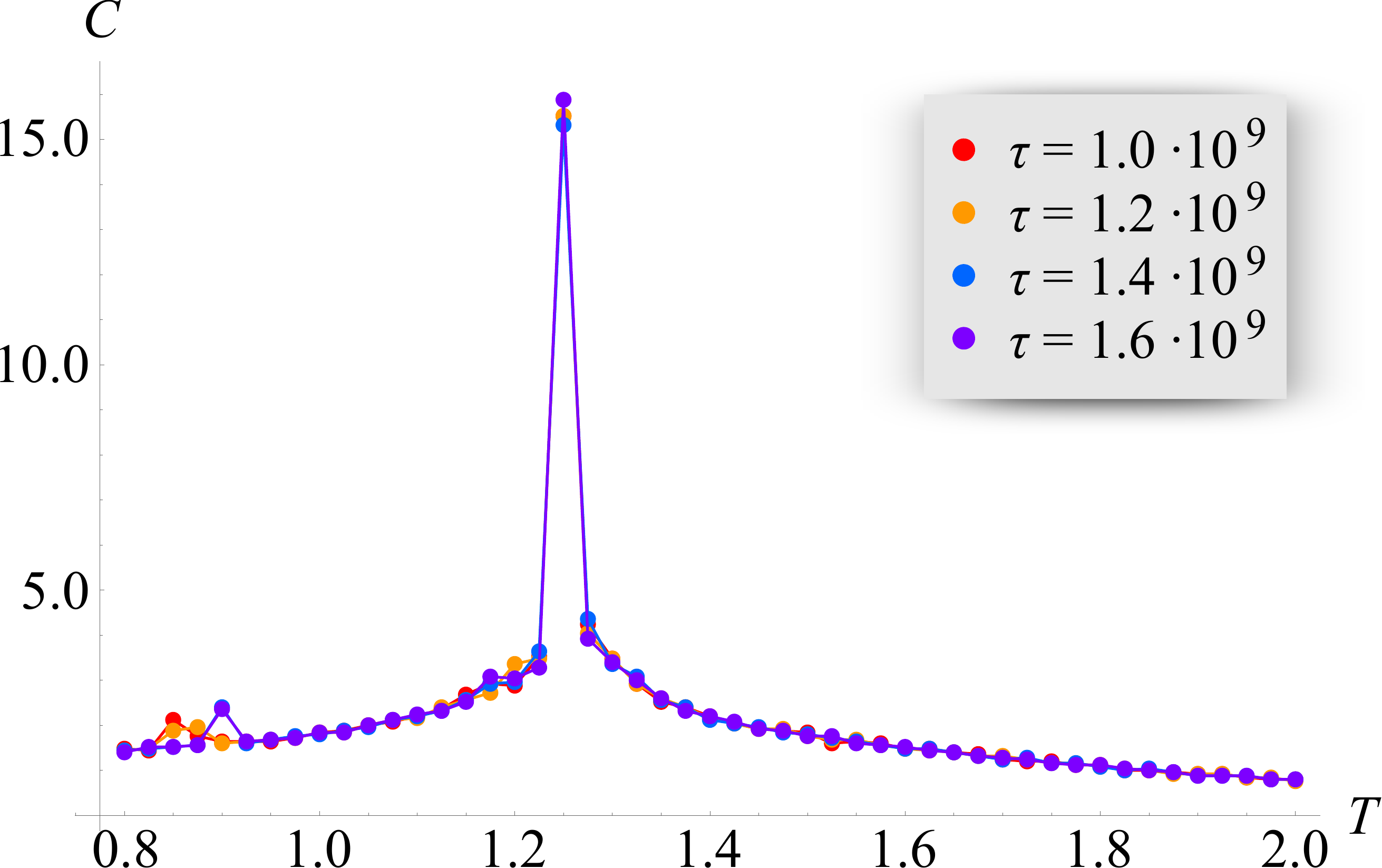}}
  \caption{\label{Cv-tau}(a) Specific heat as a function of temperature at $D=\Phi=1$ (putative 2nd order transition) for several sampling times $\tau$ given by the number of Monte Carlo iterations. (b) Specific heat at $D=0.5$, $\Phi=1$ (putative 1st order transition). In both cases $J=1$, $K=C=0$ and system size is $N=30^3$. Specific heat was calculated by taking the discrete temperature derivative $C=dE/dT$ of the internal energy $E$ per lattice site. The Monte Carlo simulations took a measurement every $5.0\cdot10^5$ iterations at each temperature; instead of dedicated annealing, temperature was swept downward in small steps between runs.}
\end{figure}

A first order phase transition may not be realized in all circumstances. Our numerical measurements at parameter combinations $(D,\Phi) \in \lbrace (1,1), (1,0.5), (0.5,0.5) \rbrace$ have so far produced only continuous changes of the order parameters and internal energy at the critical temperature $T_c$. This could be an indication of second order transitions. A comparison between the cases of putative 1st and 2nd order transitions reveals only qualitative differences. For example, the chirality FFT peak shown in Fig.\ref{Chirality}(b) decreases linearly with a fixed slope all the way to zero at $T=T_c$ in a presumed 2nd order transition, while it drops to zero with a vertical slope at $T=T_c$ in a presumed 1st order transition. The former is consistent with chirality being lost due to the continuous reduction of the thermally-averaged spin magnitude in the approach to $T_c$, while the latter is better understood via the escalating positional fluctuations of hedgehogs. Specific heat $C(T)$ plotted in Fig.\ref{Cv-tau} sends a similar message. In all cases, $C(T)$ is peaked at $T\approx T_c$, but the peak is extremely sharp in the putative 1st order transition case, its width being limited by the resolution of temperature sampling. Our attempted finite-size scaling of $C(T)$ is somewhat inconclusive as discussed in the Appendix. In the case of the putative 1st order transition, the peak width, height and location have shown negligible (below $2\%$) and non-monotonic variations with $N\in{18^3,24^3,30^3}$ and $\tau\sim 10^9$, consistent with numerical noise within effectively established thermodynamic and ergodic limits.


\section{Conclusions and Discussion}\label{secConclusions}

We studied a spin model with spin-rotation and time-reversal symmetries on the simple cubic lattice, featuring chiral interactions that can be mediated by itinerant Weyl electrons or generated from the spin-orbit coupling among localized electrons. Using classical Monte Carlo and finite-size scaling, we found that sufficiently strong isotropic chiral interactions stabilize a $4Q$ lattice of hedgehogs and antihedgehogs in the magnetically ordered state. This defect lattice is robust and present inside a large region of the phase diagram that we explored. In some cases, the hedgehog lattice melts through a first order phase transition at a finite critical temperature.

Our results are complementary to earlier studies by considering a broader set of chiral spin interactions and probing the role of positional fluctuations of topological defects in the stability of phases and the nature of phase transitions. The present study was limited to a particular portion of the phase diagram where only the $4Q$ hedgehog lattice was found. There, we observed that both the isotropic  Dzyaloshinskii-Moriya interaction and a multiple-spin isotropic chiral spin interaction contribute to the formation of the hedgehog lattice, the latter acting as a direct ``chemical potential'' for the topological defects. A more comprehensive exploration of the phase diagram will be carried out in the future, specifically in the regions with lower defect densities.

The evidence of a first order transition has important consequences. A hedgehog lattice can melt at a critical temperature and melting phase transitions are generally of the first order. The background magnetic order parameter, specifically the magnitude of the thermally-averaged local magnetization vector, is still finite just before the putative hedgehog lattice melting occurs. Therefore, it is possible that certain ferromagnetic spin correlations survive in the melted state, limited in range only by the mean separation between delocalized defects. If the spin-spin correlation length is large enough in the disordered phase near the transition temperature, then the topological defects could retain much of their identity as they now move through the system. Qualitatively, this state is a hedgehog liquid. The motion of topological defects necessarily destroys all long-range spin-spin correlations, so the question is whether this liquid state can be qualitatively distinguished from the completely featureless high-temperature phase.

It turns out that a \emph{quantum} liquid of topological defects can be a distinct thermodynamic phase. This is made possible by instanton confinement and the ensuing conservation of topological charge despite the absence of long-range order \cite{Nikolic2023a}. From a field-theory point of view \cite{Nikolic2019, Nikolic2019b}, an incompressible quantum liquid of hedgehogs can realize a chiral spin liquid with topological order and electron fractionalization -- provided that there is some imbalance between the densities of defects and antidefects. The quantum hedgehog lattice melting that would yield such a spin liquid is a first-order transition, just as the analogous quantum melting of an Abrikosov vortex lattice that produces a fractional quantum Hall state \cite{Cooper1999a, Cooper2001, Cooper2008}. However, our case is special because the densities of hedgehogs and antihedgehogs are identical. A topological order cannot be readily identified in a topologically charge-neutral liquid, so the presence of a first order transition is the main indicator that an incompressible quantum liquid, characterized by instanton confinement and probably no topological order, could be stable. In that sense, our observation of a topological melting transition in a \emph{classical} model hints at the possibility of an analogous quantum hedgehog lattice melting in the \emph{equivalent quantum model} (melting is not an inherently quantum phenomenon). Unfortunately, the Monte Carlo exploration of quantum hedgehog spin liquids is currently forbiddingly difficult for two reasons: (i) it requires simulations of equivalent four-dimensional classical spin models with string-like topological defects (hedgehog worldlines) in large space-time lattices, and (ii) it has to deal with the ``sign-problem'' which is inherent in the Berry's phase of Heisenberg quantum spins.

\section{Acknowledgements}\label{secAck}

This research was carried out at the Quantum Science and Engineering Center of George Mason University (GMU). Financial support was provided through Provost's and Physics \& Astronomy fellowships at GMU. Numerical calculations were carried out using resources provided by the Office of Research Computing at George Mason University (URL: \verb|https://orc.gmu.edu|) and funded in part by grants from the National Science Foundation (Award Number 2018631).

\appendix

\section{Metastable states and finite-size scaling}\label{secApp}

Every numerical minimization algorithm discovers local minimums without ensuring that the global minimum will be found. Monte Carlo is no exception; it is, in fact, quite prone to finding metastable states given that it samples only a small portion of the phase space. One approach in the quest for the global (free) energy minimum is to repeat Monte Carlo simulations with the same model parameters, starting from different random seed states. The hope is that different runs will land in different metastable states, one of which might be the true equilibrium state. We can decide which state might correspond to equilibrium by comparing the energy and other properties of all achieved states. Alternative approaches involve cluster updates, but we have not attempted them in the present study.

The repetition method is usually helpful in models that are not too frustrated, especially at low temperatures where Monte Carlo tends to freeze due to low probability of accepting local updates. Therefore, in anticipation of possible discontinuous transitions, one may try to anneal the system before taking measurements at any temperature. Annealing starts at a high temperature that scrambles the state and proceeds with a gradual reduction of temperature down to the desired steady value. At the end of annealing, the system is in a randomly-seeded metastable state. 

\begin{figure}[!t]
\centering
    \includegraphics[width=3.1in]{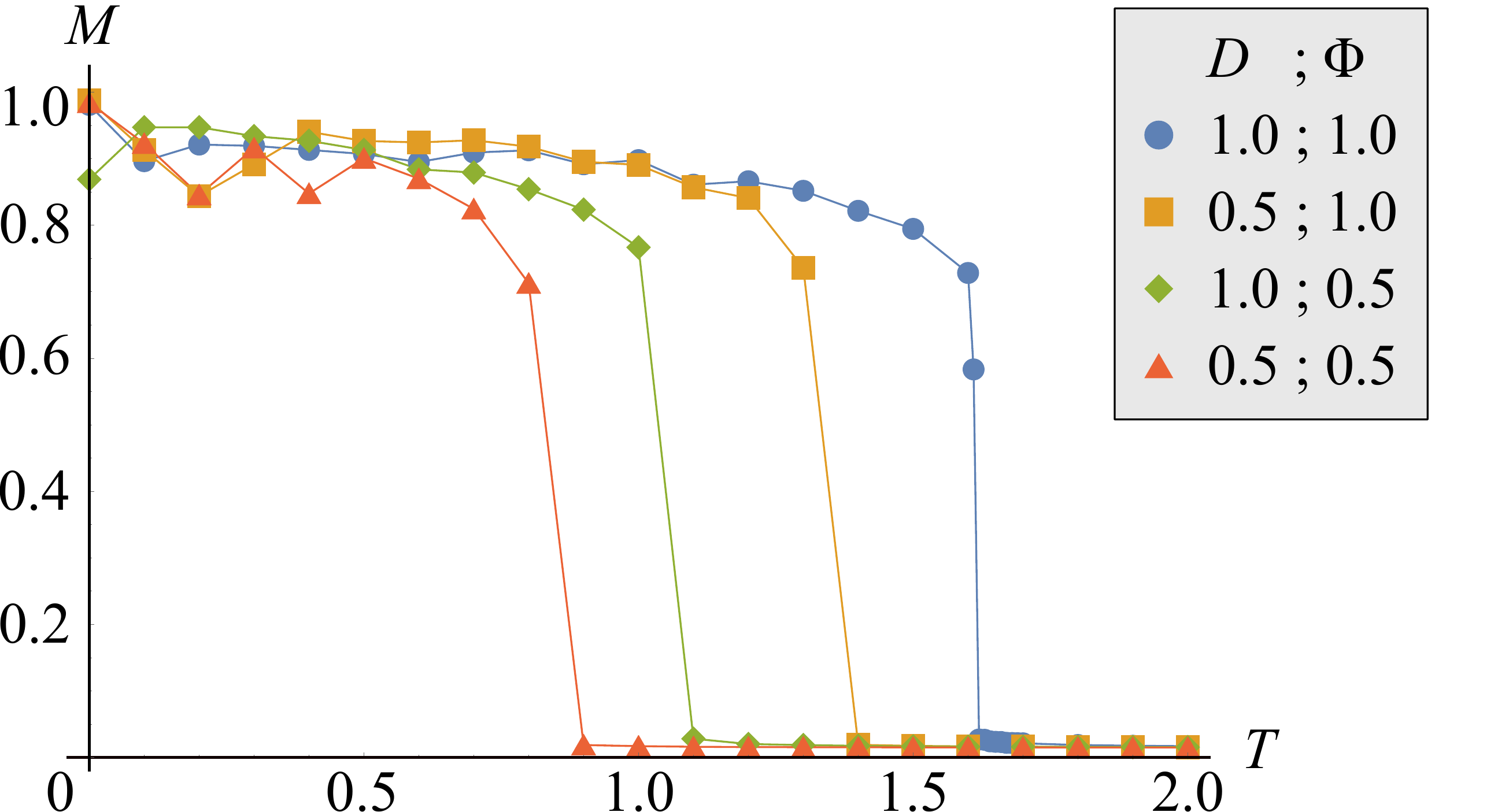}
    \caption{\label{Magnetization-meta}Thermally and lattice-averaged local magnetization magnitude (\ref{M_vs_T} as a function of temperature for various combinations of chiral couplings ($J=1$, $K=C=0$). The lattice size is $N=24^3$, and the data set for $D=\Phi=1$ includes a higher-resolution temperature sampling near $T_c$. These Monte Carlo simulations took $3000$ measurements over $7.5\times 10^8$ iterations in a steady state at every temperature. However, the employed aggressive simulated annealing for every temperature point has resulted in the discovery of the metastable $Q=\pi/2$ hedgehog lattices below the critical temperature in all runs.}
\end{figure}

We have attempted this procedure in the exploration of the thermally-driven phase transition in our model (\ref{Hamiltonian}). The extracted magnetization order parameter (\ref{M_vs_T}) tends to be noisy as a function of temperature, but exhibits a continuous drop to zero across the critical temperature if the ordered phase is a plain ferromagnet, i.e. in the absence of chiral couplings, $D=\Phi=0$. This is indeed expected of a plain classical Heisenberg model. Whenever the chiral couplings $D,\Phi\neq0$ stabilized a hedgehog lattice at low temperatures, we generally observed discontinuous changes of the magnetization magnitude, shown in Fig.\ref{Magnetization-meta}. While naively suggesting first order transitions, all simulations in this figure have discovered metastable hedgehog lattices with a different ordering wavevector and higher energy than the states reported in the main text at the same model parameters. This is attributable to annealing, which was carried out too fast.

The evidence of a first order phase transition, given in Section \ref{secPhTrans}, ultimately comes from the simulations with slow temperature sweeps that discovered the lowest energy states among many repeated runs. The results of aggressive annealing in Fig.\ref{Magnetization-meta} are tainted by metastability, but do provide a certain support for the possibility of first order transitions. Clearly, multiple hedgehog lattices exist as local minimums of the free energy across barriers from the disordered state local minimum. All hedgehog lattices are qualitatively similar. It is hard to envision a physical reason for one of these local minimums to gradually evolve out of the disordered minimum through a second order phase transition, while the other minimums, like the ones implied by Fig.\ref{Magnetization-meta}, remain apparently separated by barriers. 

We also pursued the finite-size scaling of specific heat $C(T)$ in an attempt to characterize the phase transition at $T_c$. In a second order transition, specific heat converges to $C(T)\propto |T-T_c|^{-\alpha}$ as the system size $N\to\infty$ increases, where $\alpha$ is a universal critical exponent. If $\alpha>0$, then $C(T)$ exhibits a peak that diverges at $T_c$ in the thermodynamic limit $N\to\infty$. A deceivingly similar behavior can also occur in first order phase transitions due to the latent-heat discontinuity of the internal energy density $E(T)$ picked by $C=dE/dT$.

\begin{figure}[!t]
\centering
  \subfigure[{}]{\includegraphics[width=3.5in]{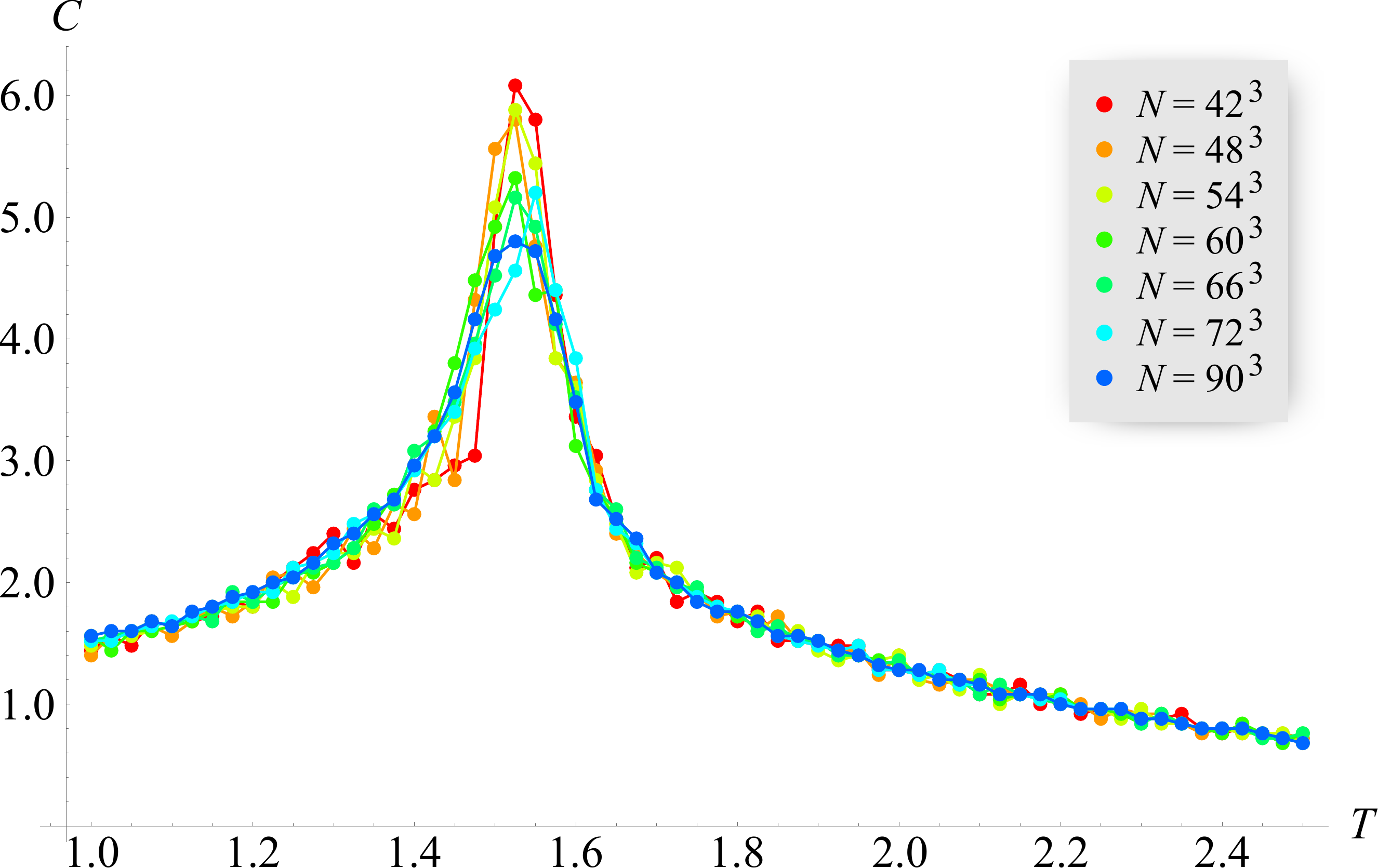}}
  \subfigure[{}]{\includegraphics[width=3.5in]{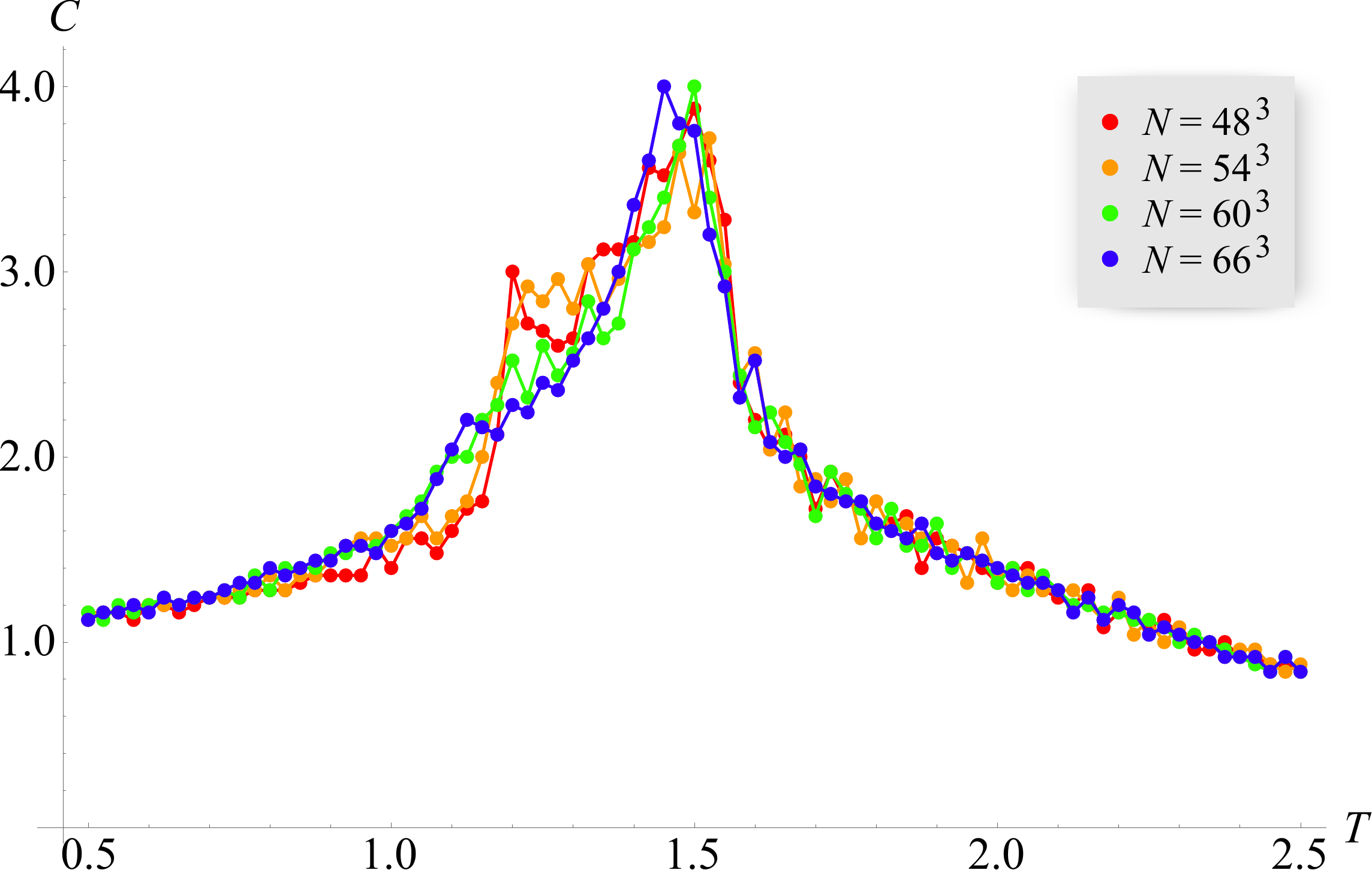}}
  \caption{\label{Cv}(a) Specific heat as a function of temperature at $D=\Phi=1$ ($J=1$, $K=C=0$), for several system sizes $N$ that are compatible with the $Q=\pi/3$ ground-state order. (b) Specific heat calculated at $D=\Phi=1$, but with a finite Kitaev-Ising coupling $K=0.5$ (the low-temperature ordered phase is still the same $Q=\pi/3$ hedgehog lattice). Specific heat was calculated by taking the discrete temperature derivative $C=dE/dT$ of the internal energy $E$ per lattice site. The Monte Carlo simulations took about $300$ measurements within $1500\times N$ iterations at each temperature; instead of dedicated annealing, temperature was swept downward in small steps between runs.}
\end{figure}

Fig.\ref{Cv} shows our calculations of $C(T)$ with two sets of model parameters that presumably produce 2nd order transitions. Limited by the available computation time, the peak height near $T\approx T_c$ is too noisy to reliably determine the peak evolution with system size (note that some system sizes are omitted as outliers). If any trend can be discerned in the panel (a), which corresponds to the example of 2nd order transition discussed in Section \ref{secPhTrans}, then the peak height appears to decrease with increasing system size $N$. This is consistent with $\alpha<0$, and perhaps related to the fact that the parent 3D Heisenberg model undergoes a 2nd order transition with $\alpha<0$ \cite{Peczak1991, Campostrini2002}. The panel (b) showcases a model with a $K\neq0$ Kitaev-Ising interaction (\ref{KitaevInteraction}), but the trend is completely inconclusive. One might naively expect to find $\alpha>0$ here because $K\neq0$ puts the model in the 3D Ising universality class \cite{Pawley1992, Balakrishnan2015}. However, the absence of a clear peak divergence suggests otherwise. It should be noted that the simulations in large systems probably failed to ergodically probe the equilibrium. This is particularly acute in the case of first order transitions, where ergodic fluctuations must cross free energy barriers that cover an exponentially large number of microstates. All of our results point to a complex landscape of metastable states separated by free energy barriers, so we speculate that our specific heat scaling becomes corrupted due to the $\tau<\tau_0$ regime explained in Section \ref{secPhTrans}. If no significant free energy barriers were present, then one would expect Monte Carlo to navigate the phase space much more easily, and perhaps reveal a critical scaling of the specific heat. In that case, we would rule out $\alpha>0$ with our results.

\begin{figure}[!t]
\centering
  \includegraphics[width=3.5in]{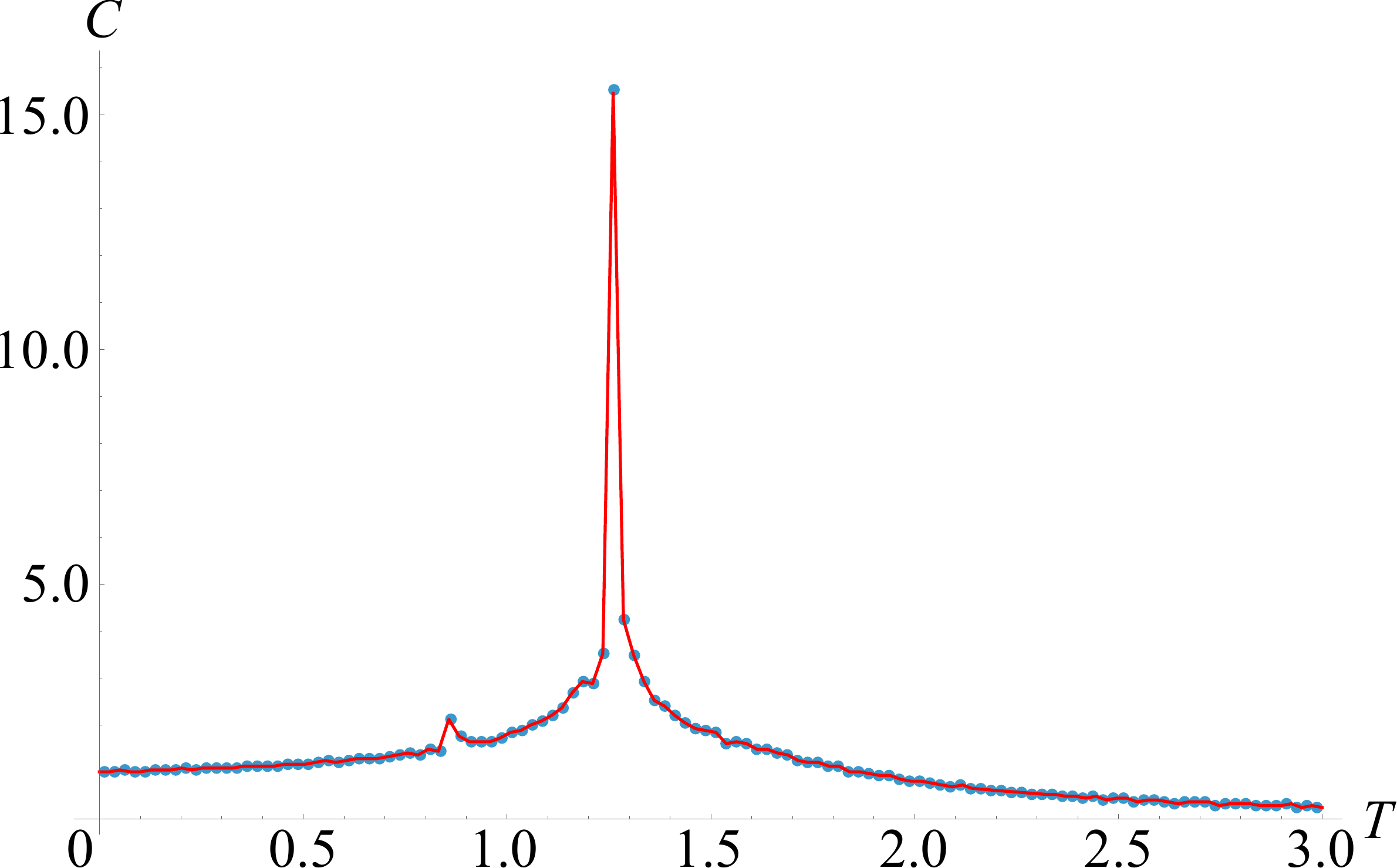}
  \caption{\label{Cv-sat}Specific heat at $D=0.5$, $\Phi=1$ in the full simulated temperature range. The data and simulation parameters are the same as in Fig.\ref{Cv-tau}(b) at $\tau=1.0\cdot10^9$.}
\end{figure}

An issue of lesser importance for our purposes is that $C(T)$ shown in Fig.\ref{Cv-sat} appears saturated at a finite value as $T\to0$. We have not taken any measures, such as cluster updates, to alleviate the freezing of Monte Carlo at low temperatures. Consequently, our $C(T)$ cannot be trusted at very lowest temperatures. That being said, it is interesting to note that $C(T)$ develops a trend toward saturation while the temperature is still relatively close ($T\sim 0.5-1$) to $T_c$ and the acceptance probability of Monte Carlo updates is appreciable ($11\%$ at $T=0.5$). Even at much lower temperatures, a nearly constant $C(T)$ implies a nearly linear internal energy $E(T)\propto T$, so that Monte Carlo keeps registering thermal fluctuations despite a very low update acceptance rate; Fig.\ref{Chirality}(b) also shows clear activity at very low temperatures, which we have interpreted as a thermal jittering of hedgehogs. Therefore, a large amount of entropy is evidently retained at lowest temperatures. This signals geometric frustration, probably associated with spin orientations inside hedgehog cores that can perhaps exhibit a macroscopic degeneracy of ground states.

\newpage



\begin{thebibliography}{46}%
\makeatletter
\providecommand \@ifxundefined [1]{%
 \@ifx{#1\undefined}
}%
\providecommand \@ifnum [1]{%
 \ifnum #1\expandafter \@firstoftwo
 \else \expandafter \@secondoftwo
 \fi
}%
\providecommand \@ifx [1]{%
 \ifx #1\expandafter \@firstoftwo
 \else \expandafter \@secondoftwo
 \fi
}%
\providecommand \natexlab [1]{#1}%
\providecommand \enquote  [1]{``#1''}%
\providecommand \bibnamefont  [1]{#1}%
\providecommand \bibfnamefont [1]{#1}%
\providecommand \citenamefont [1]{#1}%
\providecommand \href@noop [0]{\@secondoftwo}%
\providecommand \href [0]{\begingroup \@sanitize@url \@href}%
\providecommand \@href[1]{\@@startlink{#1}\@@href}%
\providecommand \@@href[1]{\endgroup#1\@@endlink}%
\providecommand \@sanitize@url [0]{\catcode `\\12\catcode `\$12\catcode
  `\&12\catcode `\#12\catcode `\^12\catcode `\_12\catcode `\%12\relax}%
\providecommand \@@startlink[1]{}%
\providecommand \@@endlink[0]{}%
\providecommand \url  [0]{\begingroup\@sanitize@url \@url }%
\providecommand \@url [1]{\endgroup\@href {#1}{\urlprefix }}%
\providecommand \urlprefix  [0]{URL }%
\providecommand \Eprint [0]{\href }%
\providecommand \doibase [0]{https://doi.org/}%
\providecommand \selectlanguage [0]{\@gobble}%
\providecommand \bibinfo  [0]{\@secondoftwo}%
\providecommand \bibfield  [0]{\@secondoftwo}%
\providecommand \translation [1]{[#1]}%
\providecommand \BibitemOpen [0]{}%
\providecommand \bibitemStop [0]{}%
\providecommand \bibitemNoStop [0]{.\EOS\space}%
\providecommand \EOS [0]{\spacefactor3000\relax}%
\providecommand \BibitemShut  [1]{\csname bibitem#1\endcsname}%
\let\auto@bib@innerbib\@empty
\bibitem [{\citenamefont {Abrikosov}(1957)}]{Abrikosov1957}%
  \BibitemOpen
  \bibfield  {author} {\bibinfo {author} {\bibfnamefont {A.~A.}\ \bibnamefont
  {Abrikosov}},\ }\bibfield  {title} {\bibinfo {title} {{On the Magnetic
  Properties of Superconductors of the Second Group}},\ }\href@noop {}
  {\bibfield  {journal} {\bibinfo  {journal} {Zhurnal Eksperimental'noi i
  Teoreticheskoi Fiziki}\ }\textbf {\bibinfo {volume} {32}},\ \bibinfo {pages}
  {1442} (\bibinfo {year} {1957})},\ \bibinfo {note} {sov. Phys. JETP {\bf 5},
  1174 (1957)}\BibitemShut {NoStop}%
\bibitem [{\citenamefont {Wang}\ \emph {et~al.}(2001)\citenamefont {Wang},
  \citenamefont {Xu}, \citenamefont {Kakeshita}, \citenamefont {Uchida},
  \citenamefont {Ono}, \citenamefont {Ando},\ and\ \citenamefont
  {Ong}}]{Ong2001}%
  \BibitemOpen
  \bibfield  {author} {\bibinfo {author} {\bibfnamefont {Y.}~\bibnamefont
  {Wang}}, \bibinfo {author} {\bibfnamefont {Z.~A.}\ \bibnamefont {Xu}},
  \bibinfo {author} {\bibfnamefont {T.}~\bibnamefont {Kakeshita}}, \bibinfo
  {author} {\bibfnamefont {S.}~\bibnamefont {Uchida}}, \bibinfo {author}
  {\bibfnamefont {S.}~\bibnamefont {Ono}}, \bibinfo {author} {\bibfnamefont
  {Y.}~\bibnamefont {Ando}},\ and\ \bibinfo {author} {\bibfnamefont {N.~P.}\
  \bibnamefont {Ong}},\ }\bibfield  {title} {\bibinfo {title} {{Onset of the
  vortexlike Nernst signal above Tc in La$_{2-x}$Sr$_x$CuO$_4$ and
  Bi$_2$Sr$_{2-y}$La$_y$CuO$_6$}},\ }\href@noop {} {\bibfield  {journal}
  {\bibinfo  {journal} {Physical Review B}\ }\textbf {\bibinfo {volume} {64}},\
  \bibinfo {pages} {224519} (\bibinfo {year} {2001})}\BibitemShut {NoStop}%
\bibitem [{\citenamefont {Wang}\ \emph {et~al.}(2006)\citenamefont {Wang},
  \citenamefont {Li},\ and\ \citenamefont {Ong}}]{Wang2006}%
  \BibitemOpen
  \bibfield  {author} {\bibinfo {author} {\bibfnamefont {Y.}~\bibnamefont
  {Wang}}, \bibinfo {author} {\bibfnamefont {L.}~\bibnamefont {Li}},\ and\
  \bibinfo {author} {\bibfnamefont {N.~P.}\ \bibnamefont {Ong}},\ }\bibfield
  {title} {\bibinfo {title} {{Nernst effect in high-$T_c$ superconductors}},\
  }\href@noop {} {\bibfield  {journal} {\bibinfo  {journal} {Physical Review
  B}\ }\textbf {\bibinfo {volume} {73}},\ \bibinfo {eid} {024510} (\bibinfo
  {year} {2006})}\BibitemShut {NoStop}%
\bibitem [{\citenamefont {Cooper}\ and\ \citenamefont
  {Wilkin}(1999)}]{Cooper1999a}%
  \BibitemOpen
  \bibfield  {author} {\bibinfo {author} {\bibfnamefont {N.~R.}\ \bibnamefont
  {Cooper}}\ and\ \bibinfo {author} {\bibfnamefont {N.~K.}\ \bibnamefont
  {Wilkin}},\ }\bibfield  {title} {\bibinfo {title} {{Composite fermion
  description of rotating Bose-Einstein condensates}},\ }\href@noop {}
  {\bibfield  {journal} {\bibinfo  {journal} {Physical Review B}\ }\textbf
  {\bibinfo {volume} {60}},\ \bibinfo {pages} {R16279} (\bibinfo {year}
  {1999})}\BibitemShut {NoStop}%
\bibitem [{\citenamefont {Cooper}\ \emph {et~al.}(2001)\citenamefont {Cooper},
  \citenamefont {Wilkin},\ and\ \citenamefont {Gunn}}]{Cooper2001}%
  \BibitemOpen
  \bibfield  {author} {\bibinfo {author} {\bibfnamefont {N.~R.}\ \bibnamefont
  {Cooper}}, \bibinfo {author} {\bibfnamefont {N.~K.}\ \bibnamefont {Wilkin}},\
  and\ \bibinfo {author} {\bibfnamefont {J.~M.~F.}\ \bibnamefont {Gunn}},\
  }\bibfield  {title} {\bibinfo {title} {{Quantum phases of vortices in
  rotating Bose-Einstein condensates}},\ }\href@noop {} {\bibfield  {journal}
  {\bibinfo  {journal} {Physical Review Letters}\ }\textbf {\bibinfo {volume}
  {87}},\ \bibinfo {pages} {120405} (\bibinfo {year} {2001})}\BibitemShut
  {NoStop}%
\bibitem [{\citenamefont {Cooper}(2008)}]{Cooper2008}%
  \BibitemOpen
  \bibfield  {author} {\bibinfo {author} {\bibfnamefont {N.~R.}\ \bibnamefont
  {Cooper}},\ }\bibfield  {title} {\bibinfo {title} {{Rapidly rotating atomic
  gases}},\ }\href@noop {} {\bibfield  {journal} {\bibinfo  {journal} {Advances
  in Physics}\ }\textbf {\bibinfo {volume} {57}},\ \bibinfo {pages} {539–616}
  (\bibinfo {year} {2008})}\BibitemShut {NoStop}%
\bibitem [{\citenamefont {Nikolić}(2020{\natexlab{a}})}]{Nikolic2019}%
  \BibitemOpen
  \bibfield  {author} {\bibinfo {author} {\bibfnamefont {P.}~\bibnamefont
  {Nikolić}},\ }\bibfield  {title} {\bibinfo {title} {{Topological orders of
  monopoles and hedgehogs: From electronic and magnetic spin-orbit coupling to
  quarks}},\ }\href@noop {} {\bibfield  {journal} {\bibinfo  {journal}
  {Physical Review B}\ }\textbf {\bibinfo {volume} {101}},\ \bibinfo {pages}
  {115144} (\bibinfo {year} {2020}{\natexlab{a}})}\BibitemShut {NoStop}%
\bibitem [{\citenamefont {Nikolić}(2020{\natexlab{b}})}]{Nikolic2019b}%
  \BibitemOpen
  \bibfield  {author} {\bibinfo {author} {\bibfnamefont {P.}~\bibnamefont
  {Nikolić}},\ }\bibfield  {title} {\bibinfo {title} {{Quantum field theory of
  topological spin dynamics}},\ }\href@noop {} {\bibfield  {journal} {\bibinfo
  {journal} {Physical Review B}\ }\textbf {\bibinfo {volume} {102}},\ \bibinfo
  {pages} {075131} (\bibinfo {year} {2020}{\natexlab{b}})}\BibitemShut
  {NoStop}%
\bibitem [{\citenamefont {Kanazawa}\ \emph {et~al.}(2011)\citenamefont
  {Kanazawa}, \citenamefont {Onose}, \citenamefont {Arima}, \citenamefont
  {Okuyama}, \citenamefont {Ohoyama}, \citenamefont {Wakimoto}, \citenamefont
  {Kakurai}, \citenamefont {Ishiwata},\ and\ \citenamefont
  {Tokura}}]{Tokura2011}%
  \BibitemOpen
  \bibfield  {author} {\bibinfo {author} {\bibfnamefont {N.}~\bibnamefont
  {Kanazawa}}, \bibinfo {author} {\bibfnamefont {Y.}~\bibnamefont {Onose}},
  \bibinfo {author} {\bibfnamefont {T.}~\bibnamefont {Arima}}, \bibinfo
  {author} {\bibfnamefont {D.}~\bibnamefont {Okuyama}}, \bibinfo {author}
  {\bibfnamefont {K.}~\bibnamefont {Ohoyama}}, \bibinfo {author} {\bibfnamefont
  {S.}~\bibnamefont {Wakimoto}}, \bibinfo {author} {\bibfnamefont
  {K.}~\bibnamefont {Kakurai}}, \bibinfo {author} {\bibfnamefont
  {S.}~\bibnamefont {Ishiwata}},\ and\ \bibinfo {author} {\bibfnamefont
  {Y.}~\bibnamefont {Tokura}},\ }\bibfield  {title} {\bibinfo {title} {{Large
  Topological Hall Effect in a Short-Period Helimagnet MnGe}},\ }\href@noop {}
  {\bibfield  {journal} {\bibinfo  {journal} {Physical Review Letters}\
  }\textbf {\bibinfo {volume} {106}},\ \bibinfo {pages} {156603} (\bibinfo
  {year} {2011})}\BibitemShut {NoStop}%
\bibitem [{\citenamefont {Kanazawa}\ \emph {et~al.}(2012)\citenamefont
  {Kanazawa}, \citenamefont {Kim}, \citenamefont {Inosov}, \citenamefont
  {White}, \citenamefont {Egetenmeyer}, \citenamefont {Gavilano}, \citenamefont
  {Ishiwata}, \citenamefont {Onose}, \citenamefont {Arima}, \citenamefont
  {Keimer},\ and\ \citenamefont {Tokura}}]{Kanazawa2012}%
  \BibitemOpen
  \bibfield  {author} {\bibinfo {author} {\bibfnamefont {N.}~\bibnamefont
  {Kanazawa}}, \bibinfo {author} {\bibfnamefont {J.-H.}\ \bibnamefont {Kim}},
  \bibinfo {author} {\bibfnamefont {D.~S.}\ \bibnamefont {Inosov}}, \bibinfo
  {author} {\bibfnamefont {J.~S.}\ \bibnamefont {White}}, \bibinfo {author}
  {\bibfnamefont {N.}~\bibnamefont {Egetenmeyer}}, \bibinfo {author}
  {\bibfnamefont {J.~L.}\ \bibnamefont {Gavilano}}, \bibinfo {author}
  {\bibfnamefont {S.}~\bibnamefont {Ishiwata}}, \bibinfo {author}
  {\bibfnamefont {Y.}~\bibnamefont {Onose}}, \bibinfo {author} {\bibfnamefont
  {T.}~\bibnamefont {Arima}}, \bibinfo {author} {\bibfnamefont
  {B.}~\bibnamefont {Keimer}},\ and\ \bibinfo {author} {\bibfnamefont
  {Y.}~\bibnamefont {Tokura}},\ }\bibfield  {title} {\bibinfo {title}
  {{Possible skyrmion-lattice ground state in the $B20$ chiral-lattice magnet
  MnGe as seen via small-angle neutron scattering}},\ }\href
  {https://doi.org/10.1103/PhysRevB.86.134425} {\bibfield  {journal} {\bibinfo
  {journal} {Physical Review B}\ }\textbf {\bibinfo {volume} {86}},\ \bibinfo
  {pages} {134425} (\bibinfo {year} {2012})}\BibitemShut {NoStop}%
\bibitem [{\citenamefont {Shiomi}\ \emph {et~al.}(2013)\citenamefont {Shiomi},
  \citenamefont {Kanazawa}, \citenamefont {Shibata}, \citenamefont {Onose},\
  and\ \citenamefont {Tokura}}]{Tokura2013b}%
  \BibitemOpen
  \bibfield  {author} {\bibinfo {author} {\bibfnamefont {Y.}~\bibnamefont
  {Shiomi}}, \bibinfo {author} {\bibfnamefont {N.}~\bibnamefont {Kanazawa}},
  \bibinfo {author} {\bibfnamefont {K.}~\bibnamefont {Shibata}}, \bibinfo
  {author} {\bibfnamefont {Y.}~\bibnamefont {Onose}},\ and\ \bibinfo {author}
  {\bibfnamefont {Y.}~\bibnamefont {Tokura}},\ }\bibfield  {title} {\bibinfo
  {title} {{Topological Nernst effect in a three-dimensional skyrmion-lattice
  phase}},\ }\href {https://doi.org/10.1103/PhysRevB.88.064409} {\bibfield
  {journal} {\bibinfo  {journal} {Physical Review B}\ }\textbf {\bibinfo
  {volume} {88}},\ \bibinfo {pages} {064409} (\bibinfo {year}
  {2013})}\BibitemShut {NoStop}%
\bibitem [{\citenamefont {Tanigaki}\ \emph {et~al.}(2015)\citenamefont
  {Tanigaki}, \citenamefont {Shibata}, \citenamefont {Kanazawa}, \citenamefont
  {Yu}, \citenamefont {Onose}, \citenamefont {Park}, \citenamefont {Shindo},\
  and\ \citenamefont {Tokura}}]{Tokura2015}%
  \BibitemOpen
  \bibfield  {author} {\bibinfo {author} {\bibfnamefont {T.}~\bibnamefont
  {Tanigaki}}, \bibinfo {author} {\bibfnamefont {K.}~\bibnamefont {Shibata}},
  \bibinfo {author} {\bibfnamefont {N.}~\bibnamefont {Kanazawa}}, \bibinfo
  {author} {\bibfnamefont {X.}~\bibnamefont {Yu}}, \bibinfo {author}
  {\bibfnamefont {Y.}~\bibnamefont {Onose}}, \bibinfo {author} {\bibfnamefont
  {H.~S.}\ \bibnamefont {Park}}, \bibinfo {author} {\bibfnamefont
  {D.}~\bibnamefont {Shindo}},\ and\ \bibinfo {author} {\bibfnamefont
  {Y.}~\bibnamefont {Tokura}},\ }\bibfield  {title} {\bibinfo {title}
  {{Real-Space Observation of Short-Period Cubic Lattice of Skyrmions in
  MnGe}},\ }\href {https://doi.org/10.1021/acs.nanolett.5b02653} {\bibfield
  {journal} {\bibinfo  {journal} {Nano Letters}\ }\textbf {\bibinfo {volume}
  {15}},\ \bibinfo {pages} {5438–5442} (\bibinfo {year} {2015})}\BibitemShut
  {NoStop}%
\bibitem [{\citenamefont {Fujishiro}\ \emph {et~al.}(2019)\citenamefont
  {Fujishiro}, \citenamefont {Kanazawa}, \citenamefont {Nakajima},
  \citenamefont {Yu}, \citenamefont {Ohishi}, \citenamefont {Kawamura},
  \citenamefont {Kakurai}, \citenamefont {Arima}, \citenamefont {Mitamura},
  \citenamefont {Miyake}, \citenamefont {Akiba}, \citenamefont {Tokunaga},
  \citenamefont {Matsuo}, \citenamefont {Kindo}, \citenamefont {Koretsune},
  \citenamefont {Arita},\ and\ \citenamefont {Tokura}}]{Fujishiro2019}%
  \BibitemOpen
  \bibfield  {author} {\bibinfo {author} {\bibfnamefont {Y.}~\bibnamefont
  {Fujishiro}}, \bibinfo {author} {\bibfnamefont {N.}~\bibnamefont {Kanazawa}},
  \bibinfo {author} {\bibfnamefont {T.}~\bibnamefont {Nakajima}}, \bibinfo
  {author} {\bibfnamefont {X.~Z.}\ \bibnamefont {Yu}}, \bibinfo {author}
  {\bibfnamefont {K.}~\bibnamefont {Ohishi}}, \bibinfo {author} {\bibfnamefont
  {Y.}~\bibnamefont {Kawamura}}, \bibinfo {author} {\bibfnamefont
  {K.}~\bibnamefont {Kakurai}}, \bibinfo {author} {\bibfnamefont
  {T.}~\bibnamefont {Arima}}, \bibinfo {author} {\bibfnamefont
  {H.}~\bibnamefont {Mitamura}}, \bibinfo {author} {\bibfnamefont
  {A.}~\bibnamefont {Miyake}}, \bibinfo {author} {\bibfnamefont
  {K.}~\bibnamefont {Akiba}}, \bibinfo {author} {\bibfnamefont
  {M.}~\bibnamefont {Tokunaga}}, \bibinfo {author} {\bibfnamefont
  {A.}~\bibnamefont {Matsuo}}, \bibinfo {author} {\bibfnamefont
  {K.}~\bibnamefont {Kindo}}, \bibinfo {author} {\bibfnamefont
  {T.}~\bibnamefont {Koretsune}}, \bibinfo {author} {\bibfnamefont
  {R.}~\bibnamefont {Arita}},\ and\ \bibinfo {author} {\bibfnamefont
  {Y.}~\bibnamefont {Tokura}},\ }\bibfield  {title} {\bibinfo {title}
  {{Topological transitions among skyrmion- and hedgehog-lattice states in
  cubic chiral magnets}},\ }\href@noop {} {\bibfield  {journal} {\bibinfo
  {journal} {Nature Communications}\ }\textbf {\bibinfo {volume} {10}},\
  \bibinfo {pages} {1059} (\bibinfo {year} {2019})}\BibitemShut {NoStop}%
\bibitem [{\citenamefont {Kanazawa}\ \emph {et~al.}(2020)\citenamefont
  {Kanazawa}, \citenamefont {Kitaori}, \citenamefont {White}, \citenamefont
  {Ukleev}, \citenamefont {Rønnow}, \citenamefont {Tsukazaki}, \citenamefont
  {Ichikawa}, \citenamefont {Kawasaki},\ and\ \citenamefont
  {Tokura}}]{Tokura2020}%
  \BibitemOpen
  \bibfield  {author} {\bibinfo {author} {\bibfnamefont {N.}~\bibnamefont
  {Kanazawa}}, \bibinfo {author} {\bibfnamefont {A.}~\bibnamefont {Kitaori}},
  \bibinfo {author} {\bibfnamefont {J.~S.}\ \bibnamefont {White}}, \bibinfo
  {author} {\bibfnamefont {V.}~\bibnamefont {Ukleev}}, \bibinfo {author}
  {\bibfnamefont {H.~M.}\ \bibnamefont {Rønnow}}, \bibinfo {author}
  {\bibfnamefont {A.}~\bibnamefont {Tsukazaki}}, \bibinfo {author}
  {\bibfnamefont {M.}~\bibnamefont {Ichikawa}}, \bibinfo {author}
  {\bibfnamefont {M.}~\bibnamefont {Kawasaki}},\ and\ \bibinfo {author}
  {\bibfnamefont {Y.}~\bibnamefont {Tokura}},\ }\bibfield  {title} {\bibinfo
  {title} {{Direct Observation of the Statics and Dynamics of Emergent Magnetic
  Monopoles in a Chiral Magnet}},\ }\href
  {https://doi.org/10.1103/PhysRevLett.125.137202} {\bibfield  {journal}
  {\bibinfo  {journal} {Physical Review Letters}\ }\textbf {\bibinfo {volume}
  {125}},\ \bibinfo {pages} {137202} (\bibinfo {year} {2020})}\BibitemShut
  {NoStop}%
\bibitem [{\citenamefont {Kitaori}\ \emph {et~al.}(2021)\citenamefont
  {Kitaori}, \citenamefont {Kanazawa}, \citenamefont {Ishizuka}, \citenamefont
  {Yokouchi}, \citenamefont {Nagaosa},\ and\ \citenamefont
  {Tokura}}]{Tokura2021}%
  \BibitemOpen
  \bibfield  {author} {\bibinfo {author} {\bibfnamefont {A.}~\bibnamefont
  {Kitaori}}, \bibinfo {author} {\bibfnamefont {N.}~\bibnamefont {Kanazawa}},
  \bibinfo {author} {\bibfnamefont {H.}~\bibnamefont {Ishizuka}}, \bibinfo
  {author} {\bibfnamefont {T.}~\bibnamefont {Yokouchi}}, \bibinfo {author}
  {\bibfnamefont {N.}~\bibnamefont {Nagaosa}},\ and\ \bibinfo {author}
  {\bibfnamefont {Y.}~\bibnamefont {Tokura}},\ }\bibfield  {title} {\bibinfo
  {title} {{Enhanced electrical magnetochiral effect by spin-hedgehog lattice
  structural transition}},\ }\href
  {https://doi.org/10.1103/physrevb.103.l220410} {\bibfield  {journal}
  {\bibinfo  {journal} {Physical Review B}\ }\textbf {\bibinfo {volume}
  {103}},\ \bibinfo {pages} {220410} (\bibinfo {year} {2021})}\BibitemShut
  {NoStop}%
\bibitem [{\citenamefont {Aji}\ \emph {et~al.}(2023)\citenamefont {Aji},
  \citenamefont {Oda}, \citenamefont {Fujishiro}, \citenamefont {Kanazawa},
  \citenamefont {Saito}, \citenamefont {Endo}, \citenamefont {Hino},
  \citenamefont {Itoh}, \citenamefont {hisa Arima}, \citenamefont {Tokura},\
  and\ \citenamefont {Nakajima}}]{Nakajima2023}%
  \BibitemOpen
  \bibfield  {author} {\bibinfo {author} {\bibfnamefont {S.}~\bibnamefont
  {Aji}}, \bibinfo {author} {\bibfnamefont {T.}~\bibnamefont {Oda}}, \bibinfo
  {author} {\bibfnamefont {Y.}~\bibnamefont {Fujishiro}}, \bibinfo {author}
  {\bibfnamefont {N.}~\bibnamefont {Kanazawa}}, \bibinfo {author}
  {\bibfnamefont {H.}~\bibnamefont {Saito}}, \bibinfo {author} {\bibfnamefont
  {H.}~\bibnamefont {Endo}}, \bibinfo {author} {\bibfnamefont {M.}~\bibnamefont
  {Hino}}, \bibinfo {author} {\bibfnamefont {S.}~\bibnamefont {Itoh}}, \bibinfo
  {author} {\bibfnamefont {T.}~\bibnamefont {hisa Arima}}, \bibinfo {author}
  {\bibfnamefont {Y.}~\bibnamefont {Tokura}},\ and\ \bibinfo {author}
  {\bibfnamefont {T.}~\bibnamefont {Nakajima}},\ }\bibfield  {title} {\bibinfo
  {title} {{Direct observations of spin fluctuations in
  hedgehog–anti-hedgehog spin lattice states in MnSi$_{1-x}$Ge$_x$ ($x=0.6$
  and $0.8$) at zero magnetic field}},\ }\href
  {https://doi.org/10.1103/physrevb.108.054445} {\bibfield  {journal} {\bibinfo
   {journal} {Physical Review B}\ }\textbf {\bibinfo {volume} {108}},\ \bibinfo
  {pages} {054445} (\bibinfo {year} {2023})}\BibitemShut {NoStop}%
\bibitem [{\citenamefont {Ishiwata}\ \emph {et~al.}(2020)\citenamefont
  {Ishiwata}, \citenamefont {Nakajima}, \citenamefont {Kim}, \citenamefont
  {Inosov}, \citenamefont {Kanazawa}, \citenamefont {White}, \citenamefont
  {Gavilano}, \citenamefont {Georgii}, \citenamefont {Seemann}, \citenamefont
  {Brandl}, \citenamefont {Manuel}, \citenamefont {Khalyavin}, \citenamefont
  {Seki}, \citenamefont {Tokunaga}, \citenamefont {Kinoshita}, \citenamefont
  {Long}, \citenamefont {Kaneko}, \citenamefont {Taguchi}, \citenamefont
  {Arima}, \citenamefont {Keimer},\ and\ \citenamefont {Tokura}}]{Tokura2020b}%
  \BibitemOpen
  \bibfield  {author} {\bibinfo {author} {\bibfnamefont {S.}~\bibnamefont
  {Ishiwata}}, \bibinfo {author} {\bibfnamefont {T.}~\bibnamefont {Nakajima}},
  \bibinfo {author} {\bibfnamefont {J.-H.}\ \bibnamefont {Kim}}, \bibinfo
  {author} {\bibfnamefont {D.~S.}\ \bibnamefont {Inosov}}, \bibinfo {author}
  {\bibfnamefont {N.}~\bibnamefont {Kanazawa}}, \bibinfo {author}
  {\bibfnamefont {J.~S.}\ \bibnamefont {White}}, \bibinfo {author}
  {\bibfnamefont {J.~L.}\ \bibnamefont {Gavilano}}, \bibinfo {author}
  {\bibfnamefont {R.}~\bibnamefont {Georgii}}, \bibinfo {author} {\bibfnamefont
  {K.~M.}\ \bibnamefont {Seemann}}, \bibinfo {author} {\bibfnamefont
  {G.}~\bibnamefont {Brandl}}, \bibinfo {author} {\bibfnamefont
  {P.}~\bibnamefont {Manuel}}, \bibinfo {author} {\bibfnamefont {D.~D.}\
  \bibnamefont {Khalyavin}}, \bibinfo {author} {\bibfnamefont {S.}~\bibnamefont
  {Seki}}, \bibinfo {author} {\bibfnamefont {Y.}~\bibnamefont {Tokunaga}},
  \bibinfo {author} {\bibfnamefont {M.}~\bibnamefont {Kinoshita}}, \bibinfo
  {author} {\bibfnamefont {Y.~W.}\ \bibnamefont {Long}}, \bibinfo {author}
  {\bibfnamefont {Y.}~\bibnamefont {Kaneko}}, \bibinfo {author} {\bibfnamefont
  {Y.}~\bibnamefont {Taguchi}}, \bibinfo {author} {\bibfnamefont
  {T.}~\bibnamefont {Arima}}, \bibinfo {author} {\bibfnamefont
  {B.}~\bibnamefont {Keimer}},\ and\ \bibinfo {author} {\bibfnamefont
  {Y.}~\bibnamefont {Tokura}},\ }\bibfield  {title} {\bibinfo {title}
  {{Emergent topological spin structures in the centrosymmetric cubic
  perovskite ${\mathrm{SrFeO}}_{3}$}},\ }\href
  {https://doi.org/10.1103/PhysRevB.101.134406} {\bibfield  {journal} {\bibinfo
   {journal} {Physical Review B}\ }\textbf {\bibinfo {volume} {101}},\ \bibinfo
  {pages} {134406} (\bibinfo {year} {2020})}\BibitemShut {NoStop}%
\bibitem [{\citenamefont {Nikolić}(2021)}]{Nikolic2020a}%
  \BibitemOpen
  \bibfield  {author} {\bibinfo {author} {\bibfnamefont {P.}~\bibnamefont
  {Nikolić}},\ }\bibfield  {title} {\bibinfo {title} {{Dynamics of local
  magnetic moments induced by itinerant Weyl electrons}},\ }\href
  {https://doi.org/10.1103/PhysRevB.103.155151} {\bibfield  {journal} {\bibinfo
   {journal} {Physical Review B}\ }\textbf {\bibinfo {volume} {103}},\ \bibinfo
  {pages} {155151} (\bibinfo {year} {2021})}\BibitemShut {NoStop}%
\bibitem [{\citenamefont {Bauer}\ \emph {et~al.}(2013)\citenamefont {Bauer},
  \citenamefont {Garst},\ and\ \citenamefont {Pfleiderer}}]{Bauer2013}%
  \BibitemOpen
  \bibfield  {author} {\bibinfo {author} {\bibfnamefont {A.}~\bibnamefont
  {Bauer}}, \bibinfo {author} {\bibfnamefont {M.}~\bibnamefont {Garst}},\ and\
  \bibinfo {author} {\bibfnamefont {C.}~\bibnamefont {Pfleiderer}},\ }\bibfield
   {title} {\bibinfo {title} {{Specific Heat of the Skyrmion Lattice Phase and
  Field-Induced Tricritical Point in MnSi}},\ }\href
  {https://doi.org/10.1103/PhysRevLett.110.177207} {\bibfield  {journal}
  {\bibinfo  {journal} {Physical Review Letters}\ }\textbf {\bibinfo {volume}
  {110}},\ \bibinfo {pages} {177207} (\bibinfo {year} {2013})}\BibitemShut
  {NoStop}%
\bibitem [{\citenamefont {Binz}\ \emph {et~al.}(2006)\citenamefont {Binz},
  \citenamefont {Vishwanath},\ and\ \citenamefont {Aji}}]{Binz2006}%
  \BibitemOpen
  \bibfield  {author} {\bibinfo {author} {\bibfnamefont {B.}~\bibnamefont
  {Binz}}, \bibinfo {author} {\bibfnamefont {A.}~\bibnamefont {Vishwanath}},\
  and\ \bibinfo {author} {\bibfnamefont {V.}~\bibnamefont {Aji}},\ }\bibfield
  {title} {\bibinfo {title} {{Theory of the Helical Spin Crystal: A Candidate
  for the Partially Ordered State of MnSi}},\ }\href
  {https://doi.org/10.1103/PhysRevLett.96.207202} {\bibfield  {journal}
  {\bibinfo  {journal} {Physical Review Letters}\ }\textbf {\bibinfo {volume}
  {96}},\ \bibinfo {pages} {207202} (\bibinfo {year} {2006})}\BibitemShut
  {NoStop}%
\bibitem [{\citenamefont {Park}\ and\ \citenamefont {Han}(2011)}]{Han2011}%
  \BibitemOpen
  \bibfield  {author} {\bibinfo {author} {\bibfnamefont {J.-H.}\ \bibnamefont
  {Park}}\ and\ \bibinfo {author} {\bibfnamefont {J.~H.}\ \bibnamefont {Han}},\
  }\bibfield  {title} {\bibinfo {title} {{Zero-temperature phases for chiral
  magnets in three dimensions}},\ }\href
  {https://doi.org/10.1103/PhysRevB.83.184406} {\bibfield  {journal} {\bibinfo
  {journal} {Physical Reviev B}\ }\textbf {\bibinfo {volume} {83}},\ \bibinfo
  {pages} {184406} (\bibinfo {year} {2011})}\BibitemShut {NoStop}%
\bibitem [{\citenamefont {Kanazawa}\ \emph {et~al.}(2016)\citenamefont
  {Kanazawa}, \citenamefont {Nii}, \citenamefont {Zhang}, \citenamefont
  {Mishchenko}, \citenamefont {Filippis}, \citenamefont {Kagawa}, \citenamefont
  {Iwasa}, \citenamefont {Nagaosa},\ and\ \citenamefont
  {Tokura}}]{Kanazawa2016}%
  \BibitemOpen
  \bibfield  {author} {\bibinfo {author} {\bibfnamefont {N.}~\bibnamefont
  {Kanazawa}}, \bibinfo {author} {\bibfnamefont {Y.}~\bibnamefont {Nii}},
  \bibinfo {author} {\bibfnamefont {X.~X.}\ \bibnamefont {Zhang}}, \bibinfo
  {author} {\bibfnamefont {A.~S.}\ \bibnamefont {Mishchenko}}, \bibinfo
  {author} {\bibfnamefont {G.~D.}\ \bibnamefont {Filippis}}, \bibinfo {author}
  {\bibfnamefont {F.}~\bibnamefont {Kagawa}}, \bibinfo {author} {\bibfnamefont
  {Y.}~\bibnamefont {Iwasa}}, \bibinfo {author} {\bibfnamefont
  {N.}~\bibnamefont {Nagaosa}},\ and\ \bibinfo {author} {\bibfnamefont
  {Y.}~\bibnamefont {Tokura}},\ }\bibfield  {title} {\bibinfo {title}
  {{Critical phenomena of emergent magnetic monopoles in a chiral magnet}},\
  }\href@noop {} {\bibfield  {journal} {\bibinfo  {journal} {Nature
  Communications}\ }\textbf {\bibinfo {volume} {7}},\ \bibinfo {pages} {11622}
  (\bibinfo {year} {2016})}\BibitemShut {NoStop}%
\bibitem [{\citenamefont {Zhang}\ \emph {et~al.}(2016)\citenamefont {Zhang},
  \citenamefont {Mishchenko}, \citenamefont {{De Filippis}},\ and\
  \citenamefont {Nagaosa}}]{Nagaosa2016}%
  \BibitemOpen
  \bibfield  {author} {\bibinfo {author} {\bibfnamefont {X.-X.}\ \bibnamefont
  {Zhang}}, \bibinfo {author} {\bibfnamefont {A.~S.}\ \bibnamefont
  {Mishchenko}}, \bibinfo {author} {\bibfnamefont {G.}~\bibnamefont {{De
  Filippis}}},\ and\ \bibinfo {author} {\bibfnamefont {N.}~\bibnamefont
  {Nagaosa}},\ }\bibfield  {title} {\bibinfo {title} {{Electric transport in
  three-dimensional skyrmion/monopole crystal}},\ }\href
  {https://doi.org/10.1103/PhysRevB.94.174428} {\bibfield  {journal} {\bibinfo
  {journal} {Physical Review B}\ }\textbf {\bibinfo {volume} {94}},\ \bibinfo
  {pages} {174428} (\bibinfo {year} {2016})}\BibitemShut {NoStop}%
\bibitem [{\citenamefont {Yang}\ \emph {et~al.}(2016)\citenamefont {Yang},
  \citenamefont {Liu},\ and\ \citenamefont {Han}}]{Han2016}%
  \BibitemOpen
  \bibfield  {author} {\bibinfo {author} {\bibfnamefont {S.-G.}\ \bibnamefont
  {Yang}}, \bibinfo {author} {\bibfnamefont {Y.-H.}\ \bibnamefont {Liu}},\ and\
  \bibinfo {author} {\bibfnamefont {J.~H.}\ \bibnamefont {Han}},\ }\bibfield
  {title} {\bibinfo {title} {{Formation of a topological monopole lattice and
  its dynamics in three-dimensional chiral magnets}},\ }\href
  {https://doi.org/10.1103/PhysRevB.94.054420} {\bibfield  {journal} {\bibinfo
  {journal} {Physical Review B}\ }\textbf {\bibinfo {volume} {94}},\ \bibinfo
  {pages} {054420} (\bibinfo {year} {2016})}\BibitemShut {NoStop}%
\bibitem [{\citenamefont {Bornemann}\ \emph {et~al.}(2019)\citenamefont
  {Bornemann}, \citenamefont {Grytsiuk}, \citenamefont {Baumeister},
  \citenamefont {dos Dias}, \citenamefont {Zeller}, \citenamefont {Lounis},\
  and\ \citenamefont {Blügel}}]{Bornemann2019}%
  \BibitemOpen
  \bibfield  {author} {\bibinfo {author} {\bibfnamefont {M.}~\bibnamefont
  {Bornemann}}, \bibinfo {author} {\bibfnamefont {S.}~\bibnamefont {Grytsiuk}},
  \bibinfo {author} {\bibfnamefont {P.~F.}\ \bibnamefont {Baumeister}},
  \bibinfo {author} {\bibfnamefont {M.~S.}\ \bibnamefont {dos Dias}}, \bibinfo
  {author} {\bibfnamefont {R.}~\bibnamefont {Zeller}}, \bibinfo {author}
  {\bibfnamefont {S.}~\bibnamefont {Lounis}},\ and\ \bibinfo {author}
  {\bibfnamefont {S.}~\bibnamefont {Blügel}},\ }\bibfield  {title} {\bibinfo
  {title} {{Complex magnetism of B20-MnGe: from spin-spirals, hedgehogs to
  monopoles}},\ }\href@noop {} {\bibfield  {journal} {\bibinfo  {journal}
  {Journal of Physics: Condensed Matter}\ }\textbf {\bibinfo {volume} {31}},\
  \bibinfo {pages} {485801} (\bibinfo {year} {2019})}\BibitemShut {NoStop}%
\bibitem [{\citenamefont {Okumura}\ \emph
  {et~al.}(2020{\natexlab{a}})\citenamefont {Okumura}, \citenamefont {Hayami},
  \citenamefont {Kato},\ and\ \citenamefont {Motome}}]{Motome2019}%
  \BibitemOpen
  \bibfield  {author} {\bibinfo {author} {\bibfnamefont {S.}~\bibnamefont
  {Okumura}}, \bibinfo {author} {\bibfnamefont {S.}~\bibnamefont {Hayami}},
  \bibinfo {author} {\bibfnamefont {Y.}~\bibnamefont {Kato}},\ and\ \bibinfo
  {author} {\bibfnamefont {Y.}~\bibnamefont {Motome}},\ }\bibfield  {title}
  {\bibinfo {title} {{Magnetic hedgehog lattices in noncentrosymmetric
  metals}},\ }\href {https://doi.org/10.1103/physrevb.101.144416} {\bibfield
  {journal} {\bibinfo  {journal} {Physical Review B}\ }\textbf {\bibinfo
  {volume} {101}},\ \bibinfo {pages} {144416} (\bibinfo {year}
  {2020}{\natexlab{a}})}\BibitemShut {NoStop}%
\bibitem [{\citenamefont {Okumura}\ \emph
  {et~al.}(2020{\natexlab{b}})\citenamefont {Okumura}, \citenamefont {Hayami},
  \citenamefont {Kato},\ and\ \citenamefont {Motome}}]{Motome2019b}%
  \BibitemOpen
  \bibfield  {author} {\bibinfo {author} {\bibfnamefont {S.}~\bibnamefont
  {Okumura}}, \bibinfo {author} {\bibfnamefont {S.}~\bibnamefont {Hayami}},
  \bibinfo {author} {\bibfnamefont {Y.}~\bibnamefont {Kato}},\ and\ \bibinfo
  {author} {\bibfnamefont {Y.}~\bibnamefont {Motome}},\ }\bibfield  {title}
  {\bibinfo {title} {{Tracing Monopoles and Anti-monopoles in a Magnetic
  Hedgehog Lattice}},\ }\bibfield  {booktitle} {\emph {\bibinfo {booktitle}
  {{Proceedings of the International Conference on Strongly Correlated Electron
  Systems (SCES2019)}}},\ }\href {https://doi.org/10.7566/jpscp.30.011010}
  {\bibfield  {journal} {\bibinfo  {journal} {JPS Conference Proceedings}\
  }\textbf {\bibinfo {volume} {30}},\ \bibinfo {pages} {011010} (\bibinfo
  {year} {2020}{\natexlab{b}})}\BibitemShut {NoStop}%
\bibitem [{\citenamefont {Aoyama}\ and\ \citenamefont
  {Kawamura}(2021)}]{Kawamura2020}%
  \BibitemOpen
  \bibfield  {author} {\bibinfo {author} {\bibfnamefont {K.}~\bibnamefont
  {Aoyama}}\ and\ \bibinfo {author} {\bibfnamefont {H.}~\bibnamefont
  {Kawamura}},\ }\bibfield  {title} {\bibinfo {title} {{Hedgehog-lattice spin
  texture in classical Heisenberg antiferromagnets on the breathing pyrochlore
  lattice}},\ }\href {https://doi.org/10.1103/physrevb.103.014406} {\bibfield
  {journal} {\bibinfo  {journal} {Physical Review B}\ }\textbf {\bibinfo
  {volume} {103}},\ \bibinfo {pages} {014406} (\bibinfo {year}
  {2021})}\BibitemShut {NoStop}%
\bibitem [{\citenamefont {Shimizu}\ \emph {et~al.}(2021)\citenamefont
  {Shimizu}, \citenamefont {Okumura}, \citenamefont {Kato},\ and\ \citenamefont
  {Motome}}]{Motome2020}%
  \BibitemOpen
  \bibfield  {author} {\bibinfo {author} {\bibfnamefont {K.}~\bibnamefont
  {Shimizu}}, \bibinfo {author} {\bibfnamefont {S.}~\bibnamefont {Okumura}},
  \bibinfo {author} {\bibfnamefont {Y.}~\bibnamefont {Kato}},\ and\ \bibinfo
  {author} {\bibfnamefont {Y.}~\bibnamefont {Motome}},\ }\bibfield  {title}
  {\bibinfo {title} {{Phase transitions between helices, vortices, and
  hedgehogs driven by spatial anisotropy in chiral magnets}},\ }\href
  {https://doi.org/10.1103/physrevb.103.054427} {\bibfield  {journal} {\bibinfo
   {journal} {Physical Review B}\ }\textbf {\bibinfo {volume} {103}},\ \bibinfo
  {pages} {054427} (\bibinfo {year} {2021})}\BibitemShut {NoStop}%
\bibitem [{\citenamefont {Kato}\ \emph {et~al.}(2021)\citenamefont {Kato},
  \citenamefont {Hayami},\ and\ \citenamefont {Motome}}]{Motome2021}%
  \BibitemOpen
  \bibfield  {author} {\bibinfo {author} {\bibfnamefont {Y.}~\bibnamefont
  {Kato}}, \bibinfo {author} {\bibfnamefont {S.}~\bibnamefont {Hayami}},\ and\
  \bibinfo {author} {\bibfnamefont {Y.}~\bibnamefont {Motome}},\ }\bibfield
  {title} {\bibinfo {title} {{Spin excitation spectra in helimagnetic states:
  Proper-screw, cycloid, vortex-crystal, and hedgehog lattices}},\ }\href
  {https://doi.org/10.1103/physrevb.104.224405} {\bibfield  {journal} {\bibinfo
   {journal} {Physical Review B}\ }\textbf {\bibinfo {volume} {104}},\ \bibinfo
  {pages} {224405} (\bibinfo {year} {2021})}\BibitemShut {NoStop}%
\bibitem [{\citenamefont {Hayami}\ and\ \citenamefont
  {Motome}(2021)}]{Motome2021a}%
  \BibitemOpen
  \bibfield  {author} {\bibinfo {author} {\bibfnamefont {S.}~\bibnamefont
  {Hayami}}\ and\ \bibinfo {author} {\bibfnamefont {Y.}~\bibnamefont
  {Motome}},\ }\bibfield  {title} {\bibinfo {title} {{Charge density waves in
  multiple-$Q$ spin states}},\ }\href
  {https://doi.org/10.1103/physrevb.104.144404} {\bibfield  {journal} {\bibinfo
   {journal} {Physical Review B}\ }\textbf {\bibinfo {volume} {104}},\ \bibinfo
  {pages} {144404} (\bibinfo {year} {2021})}\BibitemShut {NoStop}%
\bibitem [{\citenamefont {Kato}\ and\ \citenamefont
  {Motome}(2023)}]{Motome2022a}%
  \BibitemOpen
  \bibfield  {author} {\bibinfo {author} {\bibfnamefont {Y.}~\bibnamefont
  {Kato}}\ and\ \bibinfo {author} {\bibfnamefont {Y.}~\bibnamefont {Motome}},\
  }\bibfield  {title} {\bibinfo {title} {{Hidden topological transitions in
  emergent magnetic monopole lattices}},\ }\href
  {https://doi.org/10.1103/physrevb.107.094437} {\bibfield  {journal} {\bibinfo
   {journal} {Physical Review B}\ }\textbf {\bibinfo {volume} {107}},\ \bibinfo
  {pages} {094437} (\bibinfo {year} {2023})}\BibitemShut {NoStop}%
\bibitem [{\citenamefont {Shimizu}\ \emph {et~al.}(2022)\citenamefont
  {Shimizu}, \citenamefont {Okumura}, \citenamefont {Kato},\ and\ \citenamefont
  {Motome}}]{Motome2022b}%
  \BibitemOpen
  \bibfield  {author} {\bibinfo {author} {\bibfnamefont {K.}~\bibnamefont
  {Shimizu}}, \bibinfo {author} {\bibfnamefont {S.}~\bibnamefont {Okumura}},
  \bibinfo {author} {\bibfnamefont {Y.}~\bibnamefont {Kato}},\ and\ \bibinfo
  {author} {\bibfnamefont {Y.}~\bibnamefont {Motome}},\ }\bibfield  {title}
  {\bibinfo {title} {{Phase degree of freedom and topology in multiple-$Q$ spin
  textures}},\ }\href {https://doi.org/10.1103/physrevb.105.224405} {\bibfield
  {journal} {\bibinfo  {journal} {Physical Review B}\ }\textbf {\bibinfo
  {volume} {105}},\ \bibinfo {pages} {224405} (\bibinfo {year}
  {2022})}\BibitemShut {NoStop}%
\bibitem [{\citenamefont {Kato}\ and\ \citenamefont
  {Motome}(2022)}]{Motome2022c}%
  \BibitemOpen
  \bibfield  {author} {\bibinfo {author} {\bibfnamefont {Y.}~\bibnamefont
  {Kato}}\ and\ \bibinfo {author} {\bibfnamefont {Y.}~\bibnamefont {Motome}},\
  }\bibfield  {title} {\bibinfo {title} {{Magnetic field–temperature phase
  diagrams for multiple-$Q$ magnetic ordering: Exact steepest descent approach
  to long-range interacting spin systems}},\ }\href
  {https://doi.org/10.1103/PhysRevB.105.174413} {\bibfield  {journal} {\bibinfo
   {journal} {Physical Review B}\ }\textbf {\bibinfo {volume} {105}},\ \bibinfo
  {pages} {174413} (\bibinfo {year} {2022})}\BibitemShut {NoStop}%
\bibitem [{\citenamefont {Okumura}\ \emph {et~al.}(2022)\citenamefont
  {Okumura}, \citenamefont {Hayami}, \citenamefont {Kato},\ and\ \citenamefont
  {Motome}}]{Okumura2022}%
  \BibitemOpen
  \bibfield  {author} {\bibinfo {author} {\bibfnamefont {S.}~\bibnamefont
  {Okumura}}, \bibinfo {author} {\bibfnamefont {S.}~\bibnamefont {Hayami}},
  \bibinfo {author} {\bibfnamefont {Y.}~\bibnamefont {Kato}},\ and\ \bibinfo
  {author} {\bibfnamefont {Y.}~\bibnamefont {Motome}},\ }\bibfield  {title}
  {\bibinfo {title} {{Magnetic Hedgehog Lattice in a Centrosymmetric Cubic
  Metal}},\ }\href {https://doi.org/10.7566/jpsj.91.093702} {\bibfield
  {journal} {\bibinfo  {journal} {Journal of the Physical Society of Japan}\
  }\textbf {\bibinfo {volume} {91}},\ \bibinfo {pages} {093702} (\bibinfo
  {year} {2022})}\BibitemShut {NoStop}%
\bibitem [{\citenamefont {Paradezhenko}\ \emph {et~al.}(2022)\citenamefont
  {Paradezhenko}, \citenamefont {Pervishko}, \citenamefont {Swain},
  \citenamefont {Sengupta},\ and\ \citenamefont {Yudin}}]{Yudin2022}%
  \BibitemOpen
  \bibfield  {author} {\bibinfo {author} {\bibfnamefont {G.~V.}\ \bibnamefont
  {Paradezhenko}}, \bibinfo {author} {\bibfnamefont {A.~A.}\ \bibnamefont
  {Pervishko}}, \bibinfo {author} {\bibfnamefont {N.}~\bibnamefont {Swain}},
  \bibinfo {author} {\bibfnamefont {P.}~\bibnamefont {Sengupta}},\ and\
  \bibinfo {author} {\bibfnamefont {D.}~\bibnamefont {Yudin}},\ }\bibfield
  {title} {\bibinfo {title} {{Spin-hedgehog-derived electromagnetic effects in
  itinerant magnets}},\ }\href {https://doi.org/10.1039/d2cp03486g} {\bibfield
  {journal} {\bibinfo  {journal} {Physical Chemistry Chemical Physics}\
  }\textbf {\bibinfo {volume} {24}},\ \bibinfo {pages} {24317–24322}
  (\bibinfo {year} {2022})}\BibitemShut {NoStop}%
\bibitem [{\citenamefont {Eto}\ and\ \citenamefont
  {Mochizuki}(2024)}]{Mochizuki2024}%
  \BibitemOpen
  \bibfield  {author} {\bibinfo {author} {\bibfnamefont {R.}~\bibnamefont
  {Eto}}\ and\ \bibinfo {author} {\bibfnamefont {M.}~\bibnamefont
  {Mochizuki}},\ }\bibfield  {title} {\bibinfo {title} {{Theory of Collective
  Excitations in the Quadruple-$Q$ Magnetic Hedgehog Lattices}},\ }\href
  {https://doi.org/10.1103/physrevlett.132.226705} {\bibfield  {journal}
  {\bibinfo  {journal} {Physical Review Letters}\ }\textbf {\bibinfo {volume}
  {132}},\ \bibinfo {pages} {226705} (\bibinfo {year} {2024})}\BibitemShut
  {NoStop}%
\bibitem [{\citenamefont {Grytsiuk}\ \emph {et~al.}(2020)\citenamefont
  {Grytsiuk}, \citenamefont {Hanke}, \citenamefont {Hoffmann}, \citenamefont
  {Bouaziz}, \citenamefont {Gomonay}, \citenamefont {Bihlmayer}, \citenamefont
  {Lounis}, \citenamefont {Mokrousov},\ and\ \citenamefont
  {Blügel}}]{Blugel2020}%
  \BibitemOpen
  \bibfield  {author} {\bibinfo {author} {\bibfnamefont {S.}~\bibnamefont
  {Grytsiuk}}, \bibinfo {author} {\bibfnamefont {J.-P.}\ \bibnamefont {Hanke}},
  \bibinfo {author} {\bibfnamefont {M.}~\bibnamefont {Hoffmann}}, \bibinfo
  {author} {\bibfnamefont {J.}~\bibnamefont {Bouaziz}}, \bibinfo {author}
  {\bibfnamefont {O.}~\bibnamefont {Gomonay}}, \bibinfo {author} {\bibfnamefont
  {G.}~\bibnamefont {Bihlmayer}}, \bibinfo {author} {\bibfnamefont
  {S.}~\bibnamefont {Lounis}}, \bibinfo {author} {\bibfnamefont
  {Y.}~\bibnamefont {Mokrousov}},\ and\ \bibinfo {author} {\bibfnamefont
  {S.}~\bibnamefont {Blügel}},\ }\bibfield  {title} {\bibinfo {title}
  {{Topological–chiral magnetic interactions driven by emergent orbital
  magnetism}},\ }\href@noop {} {\bibfield  {journal} {\bibinfo  {journal}
  {Nature Communications}\ }\textbf {\bibinfo {volume} {11}},\ \bibinfo {pages}
  {511} (\bibinfo {year} {2020})}\BibitemShut {NoStop}%
\bibitem [{Note1()}]{Note1}%
  \BibitemOpen
  \bibinfo {note} {The DM vector is perpendicular to the lattice bonds which
  are bisected by mirror symmetry planes. However, the long range of the RKKY
  DM interactions and the reduced symmetries of the chiral magnetically ordered
  phases conspire to blunt the effect of these symmetry
  restrictions.}\BibitemShut {Stop}%
\bibitem [{\citenamefont {Sen}\ and\ \citenamefont
  {Chitra}(1995)}]{Chitra1995}%
  \BibitemOpen
  \bibfield  {author} {\bibinfo {author} {\bibfnamefont {D.}~\bibnamefont
  {Sen}}\ and\ \bibinfo {author} {\bibfnamefont {R.}~\bibnamefont {Chitra}},\
  }\bibfield  {title} {\bibinfo {title} {{Large-U limit of a Hubbard model in a
  magnetic field: Chiral spin interactions and paramagnetism}},\ }\href@noop {}
  {\bibfield  {journal} {\bibinfo  {journal} {Physical Review B}\ }\textbf
  {\bibinfo {volume} {51}},\ \bibinfo {pages} {1922} (\bibinfo {year}
  {1995})}\BibitemShut {NoStop}%
\bibitem [{\citenamefont {Lee}\ and\ \citenamefont
  {Kosterlitz}(1991)}]{Kosterlitz1991}%
  \BibitemOpen
  \bibfield  {author} {\bibinfo {author} {\bibfnamefont {J.}~\bibnamefont
  {Lee}}\ and\ \bibinfo {author} {\bibfnamefont {J.~M.}\ \bibnamefont
  {Kosterlitz}},\ }\bibfield  {title} {\bibinfo {title} {{Finite-size scaling
  and Monte Carlo simulations of first-order phase transitions}},\ }\href@noop
  {} {\bibfield  {journal} {\bibinfo  {journal} {Physical Reviev B}\ }\textbf
  {\bibinfo {volume} {43}},\ \bibinfo {pages} {3265} (\bibinfo {year}
  {1991})}\BibitemShut {NoStop}%
\bibitem [{\citenamefont {{Nikoli\ifmmode \acute{c}\else
  ć\fi{}}}(2024)}]{Nikolic2023a}%
  \BibitemOpen
  \bibfield  {author} {\bibinfo {author} {\bibfnamefont {P.}~\bibnamefont
  {{Nikoli\ifmmode \acute{c}\else ć\fi{}}}},\ }\bibfield  {title} {\bibinfo
  {title} {{Instanton confinement-deconfinement transitions: Stability of
  pseudogap phases and topological order}},\ }\href
  {https://doi.org/10.1103/PhysRevB.109.165132} {\bibfield  {journal} {\bibinfo
   {journal} {Physical Review B}\ }\textbf {\bibinfo {volume} {109}},\ \bibinfo
  {pages} {165132} (\bibinfo {year} {2024})}\BibitemShut {NoStop}%
\bibitem [{\citenamefont {Peczak}\ \emph {et~al.}(1991)\citenamefont {Peczak},
  \citenamefont {Ferrenberg},\ and\ \citenamefont {Landau}}]{Peczak1991}%
  \BibitemOpen
  \bibfield  {author} {\bibinfo {author} {\bibfnamefont {P.}~\bibnamefont
  {Peczak}}, \bibinfo {author} {\bibfnamefont {A.~M.}\ \bibnamefont
  {Ferrenberg}},\ and\ \bibinfo {author} {\bibfnamefont {D.~P.}\ \bibnamefont
  {Landau}},\ }\bibfield  {title} {\bibinfo {title} {{High-accuracy Monte Carlo
  study of the three-dimensional classical Heisenberg ferromagnet}},\ }\href
  {https://doi.org/10.1103/PhysRevB.43.6087} {\bibfield  {journal} {\bibinfo
  {journal} {Physical Review B}\ }\textbf {\bibinfo {volume} {43}},\ \bibinfo
  {pages} {6087–6093} (\bibinfo {year} {1991})}\BibitemShut {NoStop}%
\bibitem [{\citenamefont {Campostrini}\ \emph {et~al.}(2002)\citenamefont
  {Campostrini}, \citenamefont {Hasenbusch}, \citenamefont {Pelissetto},
  \citenamefont {Rossi},\ and\ \citenamefont {Vicari}}]{Campostrini2002}%
  \BibitemOpen
  \bibfield  {author} {\bibinfo {author} {\bibfnamefont {M.}~\bibnamefont
  {Campostrini}}, \bibinfo {author} {\bibfnamefont {M.}~\bibnamefont
  {Hasenbusch}}, \bibinfo {author} {\bibfnamefont {A.}~\bibnamefont
  {Pelissetto}}, \bibinfo {author} {\bibfnamefont {P.}~\bibnamefont {Rossi}},\
  and\ \bibinfo {author} {\bibfnamefont {E.}~\bibnamefont {Vicari}},\
  }\bibfield  {title} {\bibinfo {title} {{Critical exponents and equation of
  state of the three-dimensional Heisenberg universality class}},\ }\href
  {https://doi.org/10.1103/physrevb.65.144520} {\bibfield  {journal} {\bibinfo
  {journal} {Physical Review B}\ }\textbf {\bibinfo {volume} {65}},\ \bibinfo
  {pages} {144520} (\bibinfo {year} {2002})}\BibitemShut {NoStop}%
\bibitem [{\citenamefont {Baillie}\ \emph {et~al.}(1992)\citenamefont
  {Baillie}, \citenamefont {Gupta}, \citenamefont {Hawick},\ and\ \citenamefont
  {Pawley}}]{Pawley1992}%
  \BibitemOpen
  \bibfield  {author} {\bibinfo {author} {\bibfnamefont {C.~F.}\ \bibnamefont
  {Baillie}}, \bibinfo {author} {\bibfnamefont {R.}~\bibnamefont {Gupta}},
  \bibinfo {author} {\bibfnamefont {K.~A.}\ \bibnamefont {Hawick}},\ and\
  \bibinfo {author} {\bibfnamefont {G.~S.}\ \bibnamefont {Pawley}},\ }\bibfield
   {title} {\bibinfo {title} {{Monte Carlo renormalization-group study of the
  three-dimensional Ising model}},\ }\href
  {https://doi.org/10.1103/PhysRevB.45.10438} {\bibfield  {journal} {\bibinfo
  {journal} {Physical Review B}\ }\textbf {\bibinfo {volume} {45}},\ \bibinfo
  {pages} {10438–10453} (\bibinfo {year} {1992})}\BibitemShut {NoStop}%
\bibitem [{\citenamefont {Oleaga}\ \emph {et~al.}(2015)\citenamefont {Oleaga},
  \citenamefont {Salazar}, \citenamefont {{Ciomaga Hatnean}},\ and\
  \citenamefont {Balakrishnan}}]{Balakrishnan2015}%
  \BibitemOpen
  \bibfield  {author} {\bibinfo {author} {\bibfnamefont {A.}~\bibnamefont
  {Oleaga}}, \bibinfo {author} {\bibfnamefont {A.}~\bibnamefont {Salazar}},
  \bibinfo {author} {\bibfnamefont {M.}~\bibnamefont {{Ciomaga Hatnean}}},\
  and\ \bibinfo {author} {\bibfnamefont {G.}~\bibnamefont {Balakrishnan}},\
  }\bibfield  {title} {\bibinfo {title} {{Three-dimensional Ising critical
  behavior in
  ${R}_{0.6}\mathrm{S}{\mathrm{r}}_{0.4}\mathrm{Mn}{\mathrm{O}}_{3}\phantom{\rule{0.28em}{0ex}}(R=\mathrm{Pr},\phantom{\rule{0.28em}{0ex}}\mathrm{Nd})$
  manganites}},\ }\href {https://doi.org/10.1103/PhysRevB.92.024409} {\bibfield
   {journal} {\bibinfo  {journal} {Physical Review B}\ }\textbf {\bibinfo
  {volume} {92}},\ \bibinfo {pages} {024409} (\bibinfo {year}
  {2015})}\BibitemShut {NoStop}%
\end{thebibliography}

%

\end{document}